\documentclass[a4paper]{aa}
\usepackage{graphicx}
\usepackage{txfonts}
\usepackage{color}
\usepackage{natbib}
\bibpunct{(}{)}{;}{a}{}{,} 

\def\ao{Applied Optics}

\def\apjl{Astrophys. J. Letters}

\def\apjs{Astrophys. J. Suppl.}

\begin{document}

\title{Characterization of potentially habitable planets: Retrieval of atmospheric and planetary properties from emission spectra}
\titlerunning{Retrieval for potentially habitable planets}

\author{P. von Paris\inst{1,2,3} \and P. Hedelt\inst{1,2,4} \and F. Selsis\inst{1,2}  \and F. Schreier\inst{4} \and T. Trautmann\inst{4}}

\institute{Univ. Bordeaux, LAB, UMR 5804, F-33270, Floirac, France \and CNRS, LAB, UMR 5804, F-33270, Floirac, France \and Institut f\"{u}r Planetenforschung, Deutsches Zentrum f\"{u}r Luft- und Raumfahrt, Rutherfordstr. 2, 12489 Berlin, Germany \and Institut f\"{u}r Methodik der Fernerkundung, Deutsches Zentrum f\"{u}r Luft-und Raumfahrt,  Oberpfaffenhofen, 82234 We{\ss}ling, Germany}

\abstract {An increasing number of potentially habitable terrestrial planets and planet candidates are found by ongoing planet search programs. The search for atmospheric signatures to establish planetary habitability and the presence of life might be possible in the future.} {We want to quantify the accuracy of retrieved atmospheric parameters (composition, temperature, pressure) which might be obtained from infrared emission spectroscopy.} {We use synthetic observations of the atmospheres of hypothetical potentially habitable planets. These were constructed with a parametrized atmosphere model, a high-resolution radiative transfer model and a simplified noise model. The simulated observations were used to fit the model parameters. Furthermore, classic statistical tools such as $\chi ^2$ statistics and least-square fits were used to analyze the simulated observations.}
 {When adopting the design of currently planned or proposed exoplanet characterization missions, we find that emission spectroscopy could provide weak limits on surface conditions of terrestrial planets, hence their potential habitability. However, these mission designs are unlikely to allow to characterize the composition of the atmosphere of a habitable planet, even though CO$_2$ is detected. Upon increasing the signal-to-noise ratios by about a factor of 2-5 (depending on spectral resolution) compared to current mission designs, the CO$_2$ content could be characterized to within two orders of magnitude. The detection of the O$_3$ biosignature remains marginal. The atmospheric temperature structure could not be constrained. Therefore, a full atmospheric characterization seems to be beyond the capabilities of such missions when using only emission spectroscopy during secondary eclipse or target visits. Other methods such as transmission spectroscopy or orbital photometry are probably needed in order to give additional constraints and break degeneracies. }{}


\maketitle


\section{Introduction}

\begin{table*}
  \caption{Emission spectroscopy of potentially habitable planets: Science goals (species to be detected and characterized) and requirements in terms of spectral resolution and S/N ratio for two European exoplanet characterization missions. For Darwin, these S/N goals limit the number of potential targets to 40 (20) for $R$=5 ($R$=20) (see, e.g., \citealp{leger1996}). In the case of EChO, tens to hundreds of transits need to be accumulated in order to achieve the stated aims, depending on the IR magnitude of the host star \citep{tinetti2012}. Because of mission lifetime constraints, EChO is limited to M-type stars brighter than about $K$$\approx$9$^m$. Currently, there are only three transit host stars satisfying these criteria, all orbited by hot Neptunes or hot super-Earths.}\label{missions}
\begin{tabular}{l|c|cc|c}
\hline
\hline
Mission & Science requirement   & Goal resolution & Goal S/N ratio & References\\
\hline 
\hline 
EChO    & H$_2$O, CO$_2$, O$_3$  & 10         &  5             & \citet{tinetti2012},\citet{tessenyi2012} \\
Darwin  & CO$_2$, O$_3$          & 20         &  5             & \citet{cockell2009exp} \\
Darwin  & H$_2$O, CO$_2$, O$_3$   & 20         &  10            & \citet{cockell2009asbio} \\
Darwin  & CO$_2$                  & 5         &  12.5           & \citet{leger1996} \\
Darwin  & O$_3$                   & 20         &  20             & \citet{leger1996} \\
 \hline
\end{tabular}
\end{table*}

The possibility of finding potentially habitable or even inhabited terrestrial planets is one of the exciting motivations for the search of extrasolar planets. So far, more than 50 planets with (minimum) masses below ten Earth masses are known among the more than 800 detected extrasolar planets. Furthermore, recent studies based on radial-velocity planet searches claimed that the occurrence of low-mass, potentially terrestrial planets is rather large (e.g., \citealp{howard2010_occur}, \citealp{Wittenmyer2011}). For short-period orbits with periods$<$\,50 days, 11-17\,\% of all stars host at least one low-mass planet. Estimates inferred from microlensing surveys imply that the mean number of planets per star is larger than one \citep{cassan2012}.

Already, some potentially habitable (candidate) super-Earths in or very close to the habitable zone (HZ) of their central star have been discovered (\citealp{udry2007}, \citealp{mayor2009gliese},  \citealp{borucki2011}, \citealp{pepe2011}, \citealp{bonfils2012}, \citealp{anglada2012}, \citealp{delfosse2012}, \citealp{borucki2012}, \citealp{tuomi2013}). Hence, the detection of potentially habitable terrestrial planets is within reach of present-day instrumentation. The next step would be to search such potentially habitable worlds for the indications of the presence of life, so-called biomarkers. To be detectable remotely from Earth, such biomarkers are necessarily surface or atmospheric spectral signatures.

Several studies have addressed the aspect of atmospheric biosignatures (e.g., \citealp{sagan1993}, \citealp{schindler2000}, \citealp{DesMarais2002}, \citealp{selsis2002}) or the response of spectra with respect to central star and atmospheric composition (e.g.,  \citealp{Seg2003},  \citealp{Seg2005},  \citealp{kaltenegger2007}, \citealp{vasquez2013}). Further studies focused not only on the spectral response, but also on signal detectability (e.g., \citealp{Kaltenegger2009}, \citealp{deming2009}, \citealp{belu2010}, \citealp{rauer2011}, \citealp{vparis2011}). These studies calculated signal-to-noise ratios and suggested observation strategies using instrument capabilities of the planned James Webb Space Telescope (JWST). Since exoplanet characterization is not the design-driving purpose of JWST, dedicated exoplanet space mission concepts have been developed. Some of them have already been proposed to ESA or NASA. The concepts include coronographs such as SEE-COAST or ACCESS (e.g., \citealp{schneider2009}, \citealp{trauger2008}), interferometers such as Darwin (e.g., \citealp{leger1996}, \citealp{cockell2009exp}, \citealp{cockell2009asbio}) and near- to mid-IR spectrographs such as EChO or FINESSE (e.g., \citealp{tinetti2012}, \citealp{tessenyi2012}, \citealp{swain2012}). The ultimate aim of these missions is to characterize the atmospheres of (habitable) exoplanets spectroscopically. The Darwin mission, which is not pursued further by ESA, was explicitly designed for the investigation of terrestrial, potentially habitable planets. The main focus of EChO, currently in competition for the ESA M3 launch slot, is the observation of hot Jupiter and hot Neptune planets. However, an additional science opportunity is the possibility of observing (in case they would be discovered) nearby (habitable) terrestrial planets around low-mass stars \citep{tinetti2012}. Table \ref{missions} states the current mission design of Darwin and EChO  with the aimed spectral resolution and signal-to-noise (S/N) ratio for reaching the scientific goals and opportunities regarding terrestrial planets.

However, the difficulty of characterizing the atmospheres of small, potentially low-mass exoplanets even with high spectral resolution and good S/N ratios is illustrated by the case of GJ\,1214\,b. Numerous studies using both ground- and space-based spectroscopy and spectrophotometry (e.g, \citealp{bean2010}, \citealp{desert2011}, \citealp{croll2011gj1214}, \citealp{crossfield2011}, \citealp{bean2011}, \citealp{mooij2012}, \citealp{berta2012}) aimed at constraining the range of possible atmospheric scenarios. Most observations favor a cloud-free water vapor atmosphere. However, a data point presented by \citet{mooij2012} seems to favor a hydrogen-rich atmosphere with a possible low-altitude cloud layer. 

In view of such difficulties, several methods have been proposed recently to allow for an efficient search of sets of atmospheric parameters which produce a reasonable fit to the data. Studies by, e.g., \citet{madhusudhan2009}, \citet{lee2012} or  \citet{line2012} focused on hot Jupiter planets, with both primary and secondary eclipse data from real observations, whereas \citet{benneke2012} investigated in detail the potential of transmission spectroscopy to characterize the atmospheres of hypothetical super-Earths. These studies have shown that it is in principle feasible to constrain atmospheric composition and temperature structure for hot Jupiters and super-Earths, however given the current data quality, both in terms of spectral coverage and signal-to-noise ratios, error bars on retrieved atmospheric parameters are rather large. 

Such retrieval studies could provide valuable clues to the design of the aforementioned space missions and should be taken into account when calculating integration times or allocating observations. Therefore, we present here for the first time a study of the possible retrieval of atmospheric and planetary parameters for hypothetical habitable terrestrial planets from emission spectra. The focus of the study will be on possible constraints on surface conditions and detection of important molecules in the context of life, such as ozone and water. In contrast to the work by, e.g., \citet{lee2012} or \citet{benneke2012}, we do not use here a Bayesian optimal estimation approach but rather focus on straightforward $\chi^2$ calculations and nonlinear least-squares fits to illustrate the difficulties of atmospheric retrieval for habitable planets. From there, we investigate quantitatively the potential of planned or near-future space instrumentation to characterize habitable planets.

The paper is organized as follows: Section \ref{modeldescr}
presents atmospheric scenarios and models used. Results will be
shown in Sect. \ref{resultsect} and discussed in Sect. \ref{discussion}. Conclusions are given in Sect.
\ref{concl}.

\section{Models and atmospheric scenarios}

\label{modeldescr}

\subsection{Parametric atmosphere model}

\label{theatmosphere}

We use a semi-empirical parametric atmosphere model to describe the temperature structure and chemical composition of an arbitrary terrestrial, cloud-free habitable atmosphere, similar to the approach presented in \citet{madhusudhan2009}, \citet{line2012} or \citet{lee2012}. Note that simple, analytical 1D models have been proposed recently \citep{robinson2012} which might also be applicable to the retrieval problem.

The model atmosphere is divided into 50 layers, with a higher resolution in $\log(p)$ in the lower atmosphere where the temperature gradient is expected to be larger. We make the following assumptions regarding our temperature structure:

\begin{itemize}
 \item The lower atmosphere is convective, i.e. a troposphere exists. The temperature gradient is given by the wet adiabatic lapse rate, taking into account the release of latent heat by condensation of water:
  
\begin{equation}
\label{lapse}
\frac{d\ln p}{d\ln T}=\Gamma_{\rm{wet}}=\Gamma_{\rm{dry}}\cdot \frac{(1+\frac{LM_w}{RT}c_{\rm{H_2O}})(1+fc_{\rm{H_2O}})}{1+\frac{fM_wL^2}{c_pRT^2}c_{\rm{H_2O}}}
\end{equation}
where $\Gamma_{\rm{dry}}=\frac{c_p}{R_{\rm{gas}}}$ ($c_p$ heat capacity of the atmosphere, $R_{\rm{gas}}$ the universal gas constant) is the dry adiabatic lapse rate in pressure coordinates, $c_{\rm{H_2O}}$ is the concentration of water, $M_w$ the molar mass of water, $L$ the latent heat and $f=\frac{M}{M_w}$ the mass ratio between dry air and water. Throughout this work, we took $L$=583\,cal g$^{-1}$, i.e. its value at 300\,K.

  \item Above the tropopause, the atmosphere is in radiative equilibrium.  The temperature gradient is assumed to be linear in $\ln p$, given by a linear interpolation between the tropopause temperature and the temperature at the model lid.

\end{itemize}

\begin{figure}[h]
  \resizebox{\hsize}{!}{\includegraphics*{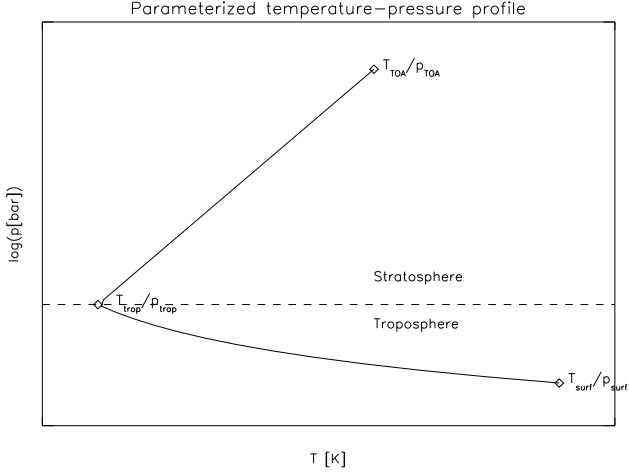}}\\
\caption{Illustration of the parameterized temperature structure.}
  \label{t_illu}
\end{figure}

Four parameters are used for the temperature-pressure profile $T=T(p)$. These are two temperatures, $T_{\rm{TOA}}$ (temperature at the top of the atmosphere, TOA) and $T_{\rm{surf}}$ (surface temperature) as well as two pressures, $p_{\rm{trop}}$ (location of the tropopause) and $p_{\rm{dry}}$ (dry surface pressure). The total surface pressure $p_{\rm{surf}}$ is constructed from $p_{\rm{dry}}$ and the water vapor pressure $p_{\rm{vap,H_2O}}(T_{\rm{surf}})$ at the specified surface temperature $T_{\rm{surf}}$, i.e. $p_{\rm{surf}}(T_{\rm{surf}})=p_{\rm{dry}}+p_{\rm{vap,H_2O}}(T_{\rm{surf}})$. The TOA pressure $p_{\rm{TOA}}$ is fixed at $p_{\rm{TOA}}$=$10^{-4}$\,bar since the low- to medium-resolution emission spectra presented in this work are not likely to be sensitive to pressures much lower than that. Fig. \ref{t_illu} illustrates our parametric temperature-pressure profile.

For $p_{\rm{trop}}$$\leqslant$$p$$\leqslant$$p_{\rm{surf}}$, the temperature $T^{(n)}$ in a layer $n$ (starting at the layer above the surface, with $n_{\rm{surface}}$=50) is calculated via (see eq. \ref{lapse})

\begin{equation}
\label{lower_atmo}
T^{(n)}=T^{(\rm{n+1})}\cdot \exp\left(-\frac{\ln \left(\frac{p^{(\rm{n+1})}}{p^{(\rm{n})}}\right)}{\Gamma_{\rm{wet}}}\right)
\end{equation}

For $p_{\rm{TOA}}$$\leqslant$$p$$\leqslant$$p_{\rm{trop}}$, the temperature profile is calculated with

\begin{equation}
\label{upper_atmo}
T^{(n)}=T_{\rm{TOA}}+\frac{T_{\rm{trop}}-T_{\rm{TOA}}}{\ln\left(\frac{p_{\rm{trop}}}{p_{\rm{TOA}}}\right)} \cdot \ln\left(\frac{p^{(n)}}{p_{\rm{TOA}}}\right)
\end{equation}
where $T_{\rm{trop}}$ (the temperature of the tropopause at $p_{\rm{trop}}$) is calculated based on the values of $T_{\rm{surf}}$, $p_{\rm{surf}}$ and $p_{\rm{trop}}$ (see eq. \ref{lower_atmo}).

The atmosphere is assumed to be composed of four species, i.e. CO$_2$, H$_2$O, the so-called biomarker O$_3$ as well as N$_2$. We chose CO$_2$ and H$_2$O because they are the main greenhouse gases on Venus, Earth and Mars, and both are generally accepted constituents of the atmospheres of potentially habitable planets. N$_{2}$ is a major atmospheric species for all terrestrial-like atmospheres in the solar system (i.e., Venus, Earth, Mars, Titan). We chose O$_{3}$ as the biomarker in our model atmospheres because it is usually considered to be a very prominent bioindicator (e.g., \citealp{sagan1993}, \citealp{schindler2000}, \citealp{selsis2002}), due to its strong mid-IR absorption band centered at 9.6\,$\mu$m. On Earth, O$_{3}$ is an indirect biomarker because its presence is due to the large O$_{2}$ amount in the atmosphere, which itself is produced by the biosphere. Note that our model atmospheres do not contain O$_2$ which would affect the temperature structure in our model mainly through its contribution to the heat capacity (Eq. \ref{lapse}) and the spectrum by a pressure broadening effect on spectral lines and potential collision-induced absorption, similar to N$_2$, but the effect would be rather small.

 The following assumptions are used for the atmospheric composition:

\begin{itemize}
 \item CO$_2$, O$_3$ and N$_2$ are assumed to follow isoprofiles. While CO$_2$ and N$_2$ are indeed isoprofiles (at least on Earth, Venus and Mars, where neither species significantly condenses), assuming an isoprofile for O$_3$ is a simplifying assumption made in this work. Because O$_3$ is formed by a 3-body reaction that requires both a pressure which is large enough and atomic oxygen produced by UV photolysis at low pressure levels, its abundance exhibits a very distinct maximum at mid-altitudes, as confirmed by many photochemical model studies. The terrestrial O$_{3}$ profile can be approximated by a Gaussian profile without taking into account interactive photochemistry (e.g., \citealp{vparis2011}). Instead of one parameter for the O$_{3}$ isoprofile, this would introduce at least three parameters into the model, namely the location of the maximum, the concentration at the maximum and the full width at half maximum of the Gaussian which would complicate retrieval further (see also discussion of the effect on spectra in Sect. \ref{atmspec}).

  \item H$_2$O and O$_3$ are minor species, i.e. are only present at a level of several $10^{-2}$ or less, hence are neglected in the calculation of the heat capacity or the mean molecular weight of the atmosphere.
  \item  Up to $p_{\rm{trop}}$, H$_2$O is calculated based on a fixed, constant relative humidity $h$, i.e. the partial pressure of water is obtained from  $p_{\rm{H_2O}}(T)=h\cdot p_{\rm{vap,H_2O}}(T)$, with the saturation vapor pressure according to local temperature. At $p_{\rm{trop}}$, we assume a cold trap, i.e. for pressures lower than $p_{\rm{trop}}$, H$_2$O remains at the value calculated for $p_{\rm{trop}}$. Using a constant relative humidity $h$ in our model is also a simple assumption. On Earth, $h$ is a function of altitude, longitude, latitude and highly variable (standard globally-averaging 1D models often use, e.g., \citealp{manabewetherald1967}, which decreases monotonically with altitude). However, since the hydrological cycle on Earth is already difficult to model correctly, it will be much harder on exoplanets where no topographic or other constraints are available. Therefore, assuming a fixed, constant relative humidity in a first step is justified.
  \item N$_2$ is a filling gas, i.e. its concentration is adjusted such that the sum over all species is unity at the surface. 
\end{itemize}

Hence, the four atmospheric constituents are fully described by the temperature profile and three additional parameters ($h$, CO$_2$ and O$_3$).

Note that the possibility to constrain surface conditions and molecular absorption bands could be limited substantially by the presence of clouds (e.g., \citealp{tinetti2006synth}, \citealp{kitzmann2011}). In that sense, the cloud-free model used in this work illustrates upper limits on detectability.

In order to calculate the heat capacity in eq. \ref{lapse}, we used the respective values for N$_2$ and CO$_2$, taken at 300\,K. The values are (in units of cal\,K$^{-1}$\,mol$^{-1}$) , $c_{\rm{p,N_2}}=6.953$ and $c_{\rm{p,CO_2}}=9.215$.

The atmospheric altitude profile is then constructed by assuming hydrostatic equilibrium, given the mean atmospheric weight, temperature-pressure profile and surface gravity. The surface gravity is calculated from the planetary mass and radius ($m_P$ and $r_P$, respectively). At a given value of $m_P$, $r_P$ is obtained from a mass-radius relationship for rocky planets of \citet{sotin2007} ($M_{\oplus}$, $R_{\oplus}$ Earth mass and radius, respectively).

\begin{equation}
\label{mrr}
\frac{r_P}{R_{\oplus}}=\left( \frac{m_P}{M_{\oplus}}\right)^{0.274}
\end{equation}

All of the 8 parameters (4 T/p, 3 chemical, 1 planetary) are uncorrelated in our model. In reality, this is not the case. For example, the surface temperature is of course related to surface pressure and atmospheric composition via the greenhouse effect. In turn, the chemical composition is influenced by atmospheric temperature through equilibrium and photochemistry and temperature-dependent surface processes (e.g., outgassing, weathering). However, for the purpose of an illustration of a retrieval algorithm, a (self-) consistent atmospheric modeling would be computationally too expensive. Self-consistency is interesting when it allows to decrease the number of parameters. In the case of cool planetary atmospheres considered here, that are far from the thermodynamic equilibrium, self-consistency is out of reach: it implies, among other things, 3D time-dependent modeling with a realistic treatment of clouds and topography. Not only this cannot be used for retrieval but in practice it also increases the number of model parameters.

\subsection{Target scenario}

\label{target}

For our analysis, we used modern Earth as the reference case for an inhabited planet. We chose model parameters as to approximately reproduce the modern Earth temperature profile (see Table \ref{atmomodels}). The relative humidity was set to $h$=0.5 throughout the troposphere, CO$_2$ to $3.55 \times10^{-4}$ and O$_3$ to $10^{-7}$. Note that an O$_3$ concentration of $3.75 \times10^{-7}$ would reproduce the atmospheric column amount of modern Earth. The planetary mass $m_P$ was fixed at 1\,$M_{\oplus}$ (hence, the model planet has a radius of 1\,$R_{\oplus}$, see Eq. \ref{mrr}).

\begin{table}[h]
  \caption{Target model parameters used in this work}\label{atmomodels}
\begin{tabular}{ll}
\hline
\hline
 $T_{\rm{TOA}}$ [K] & 270 \\
$T_{\rm{surf}}$ [K] & 290 \\
$p_{\rm{trop}}$ [bar]& 0.3 \\
$p_{\rm{dry}}$ [bar]& 1.0 \\
humidity&    0.5 \\
CO$_2$ vmr & $3.55 \times10^{-4}$\\
 O$_3$ vmr&   $10^{-7}$ \\
$m_P$ [$M_{\oplus}$]&1.0 \\
\hline
\hline 
\end{tabular}
\end{table}

\begin{figure}[h]
  \resizebox{\hsize}{!}{\includegraphics*{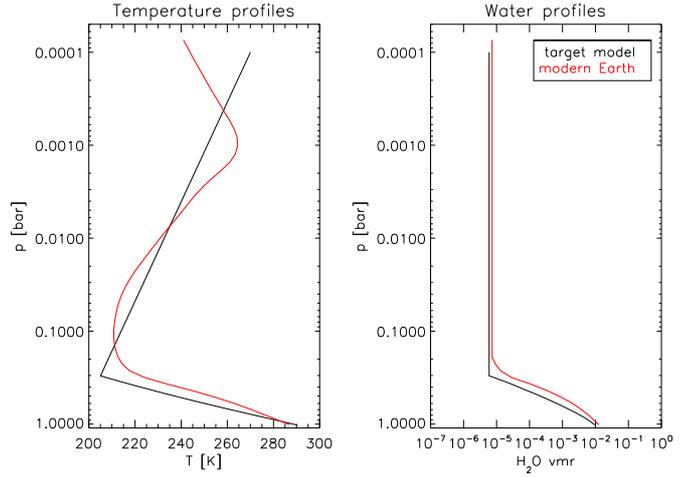}}\\
\caption{Temperature and water profiles of the target scenario, in comparison to modern Earth.}
  \label{model_tp}
\end{figure}

Fig. \ref{model_tp} shows the temperature-pressure and water-pressure profiles in comparison with a modern Earth reference case. This reference case was obtained with a consistent 1D radiative-convective climate model, coupled to photochemistry (for a detailed description, see \citealp{rauer2011} or \citealp{grenfell2011}).

Despite the simplicity of the model, the general agreement between the target profile and the modern Earth reference is rather good. Also, the water profiles agree qualitatively well, even though the target profile is calculated with a constant relative humidity.

\subsection{Planetary spectra}

\label{atmspec}

The line-by-line code MIRART-SQuIRRL \citep{schreier2003} was used to calculate high-resolution synthetic radiance spectra of the model atmospheres described above. Line parameters were taken from the Hitran 2008 database \citep{rothman2009}. The zenith angle was fixed at 38$^{\circ}$. To obtain disk-integrated emission spectra, the radiance spectra were multiplied by $\pi$. For the purpose of this work, collision-induced self-continua of CO$_2$ \citep{wordsworth2010cont} and N$_2$ (\citealp{borysow1986n2n2}, \citealp{lafferty1996}) have been included.

\begin{figure}[h]
\resizebox{\hsize}{!}{  \includegraphics*{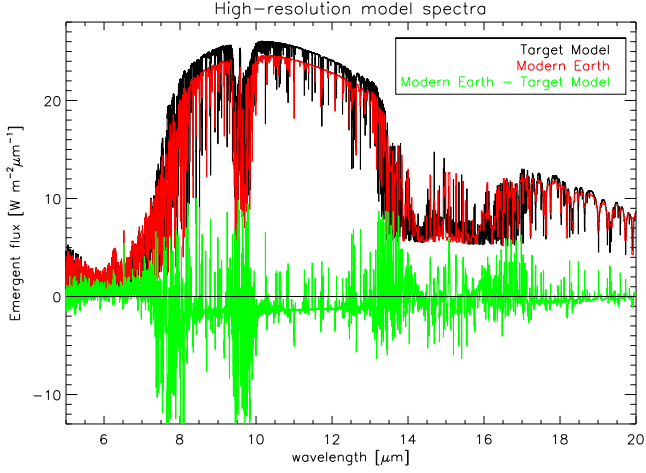}}\\
\caption{High-resolution emergent spectrum for the target model of Table \ref{atmomodels} and a comparison with modern Earth (taken from \citealp{grenfell2011})}.
  \label{model_highres}
\end{figure}

The resulting high-resolution emission spectrum for our target model of Table \ref{atmomodels} is shown in Fig. \ref{model_highres}. It is compared to a spectrum obtained using global mean modern Earth profiles (see Sect. \ref{target}). The water features (around the 6.3\,$\mu$m band and in the rotation band longwards of around 16\,$\mu$m) do not differ by much, indicating that the choice of relative humidity is indeed not critical. Some noticeable differences, however, occur in the optically thin window region between 8 and 12\,$\mu$m (due to different surface temperatures, modern Earth 288\,K, our target scenario 290\,K and the different O$_3$ profile) and around 7.7\,$\mu$m (due to a strong methane absorption band, with methane not being included in our model atmospheres). Near the 15\,$\mu$m CO$_2$ band, some differences in the spectra are apparent which are due to the slightly different temperature structure in the stratosphere (see Fig. \ref{model_tp}).

For modern Earth, the O$_3$ band probes the upper troposphere and lower stratosphere, at pressures of about 10$^{-2}$ to 10$^{-1}$ bar, where temperatures are increasing with altitude. Hence, an isoprofile which would reproduce the atmospheric column amount of O$_3$ results in less emission from the O$_3$ band since the contribution function would peak closer to the tropopause. This would imply that assuming an O$_3$ isoprofile would underestimate the emission of the O$_3$ band. However, many atmospheric modeling studies have shown that, e.g., planets orbiting around M and K stars would not show a strong stratospheric inversion (e.g., \citealp{Seg2003}, \citealp{Grenf2007asbio}). In these cases, an O$_3$ isoprofile would actually overestimate the O$_3$ band. Therefore, since we do not include the influence of the central star in our model (see above), assuming an O$_3$ isoprofile is justified for the sake of model simplicity.

\subsection{Observational model}

\label{obsmodel}

The measured quantity for spectral observations is the number of photons in a wavelength bin. To translate the calculated high-resolution spectra from Sect. \ref{atmspec} into a number of photons, we place the hypothetical target planet at a distance of 10\,pc. We furthermore assume an ideal detector (i.e., throughput and quantum efficiency of unity) with a 10\,m$^2$ collecting surface (i.e., about 3.6\,m circular telescope aperture) and an integration time of 1 hour. Then, the spectra were  binned to equidistant bins, corresponding to different spectral resolutions ($R$=5-100) at 10\,$\mu$m (i.e., bin sizes of 2-0.1\,$\mu$m), over a spectral range of 5-20\,$\mu$m, thus yielding a noise-free spectrum $I_F$. 

When comparing observed spectra and models to estimate atmospheric or planetary properties, it is important to consider the signal-to-noise (S/N) ratio per bin of the observations. We take the noise source $N_S$ to be constant over wavelength. This is of course an approximation, since many individual, wavelength-dependent terms contribute to the overall noise (e.g., photon noise of both star and planet, thermal noise of the telescope and instrument, zodiacal light etc.). However, this approach yields a simple tool to calculate synthetic observations independent of telescope and instrument configurations. We define $N_S$ as

\begin{equation}\label{noise_def}
N_S(\lambda)=\frac{B_{288}(\lambda_0)}{f_N}=\rm{const.}
\end{equation}
where $B_{288}$ is a blackbody emitter of one Earth radius ($R_{\oplus}$) with a temperature of 288\,K at a distance of 10\,pc,  $\lambda_0$= 10\,$\mu$m and $f_N$ an arbitrary noise factor.

 In order to construct noisy spectra $I_N$, we added a Gaussian noise of amplitude $A_N=\sqrt{N_S}$ to $I_F$ :

\begin{equation}\label{gauss}
I_{N,i}=I_{F,i}+\mathcal{R}_i\cdot A_N
\end{equation}
where $i$ is the wavelength bin and $\mathcal{R}_i$ drawn randomly from the standard normal distribution $\mathcal{N}_{0,1}$. The S/N ratio per bin $i$ is then

\begin{equation}\label{snr_def}
(S/N)_i=\frac{I_{N,i}}{A_N}
\end{equation}

\begin{figure}[h]
  \resizebox{\hsize}{!}{\includegraphics*{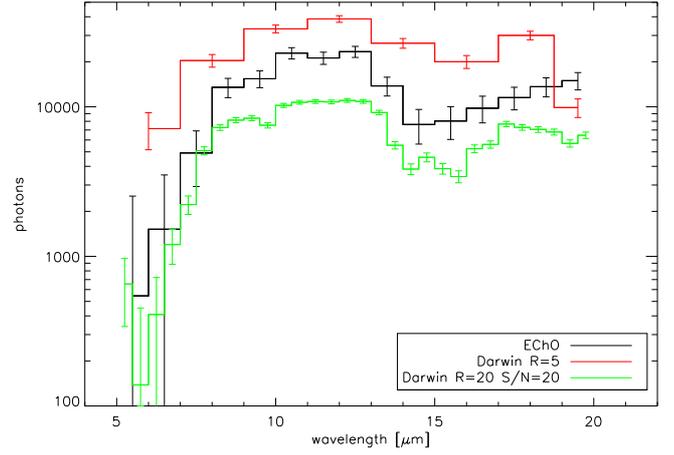}}\\
\caption{Example spectra of the target model (see Table \ref{atmomodels} and Fig. \ref{model_highres}) for different mission designs as summarized in Table \ref{missions}.}
  \label{samplespec}
\end{figure}

In order to reproduce the EChO and Darwin S/N ratio specifications at the respective spectral resolution (see Table \ref{missions}), we adapted the noise factor $f_N$ in eq. \ref{noise_def} accordingly (e.g., for EChO, $f_N$=5$\cdot$10$^{-3}$ at $R$=10 to produce a mean S/N per bin of 5-6). Fig. \ref{samplespec} shows examples of the thus generated synthetic spectra.

\subsection{Fit models}

\label{fitmodel}

When analyzing (noisy) observations and comparing them to a model, one standard quantity which is usually evaluated is the sum of the weighted squared residuals, the so-called $\chi^2$. The $\chi^2$ is defined as

\begin{equation}
\label{chi_def}
\chi^2(\vec{x})=\sum_{i=1}^N \left(\frac{O_i-M_i(\vec{x})}{\sigma_i}\right)^2
\end{equation}
with $\vec{x}$ the vector of parameters (in our case, 8 parameters, see Sect. \ref{theatmosphere}), $O_i$ the noisy observation, $M_i$ the parameter-dependent model value and $\sigma_i$ the noise level (obtained from eq. \ref{noise_def}\textbf{, i.e. $\sigma_i=A_N=\sqrt{N_S}$}) in a respective spectral bin $i$ and $N$ is the number of data points.

The $\chi^2$ as defined above in eq. \ref{chi_def} is the log-likelihood in the case of Gaussian measurement errors, which we assume here. The $n \sigma$ uncertainties on retrieved model parameters are then calculated by $\chi^{2}-\chi^{2}_{min} \leqslant n^{2}$, e.g. a 3\,$\sigma$ uncertainty corresponds to a distance of  $\Delta \chi^2$=9 from the minimum $\chi^{2}$ value ($\chi^{2}_{min}$). Finding a best-fit model corresponding to a certain minimum $\chi^{2}_{min}$, however, does not guarantee a good fit corresponding to the global minimum. Two quantities are used generally to check the quality of a fit, i.e. the $p$ value and the reduced $\chi^{2}_{\rm{red}}$. The $p$ value is the probability of generating a worse (i.e., larger) $\chi^2$ than the one calculated if the chosen parameters were the true ones. All parameter vectors $\vec{x}$ with $\chi^2(\vec{x})$$<$$\chi^2_p$ would be considered a good fit, with $\chi^2_p$ the $\chi^2$ value corresponding to the chosen $p$. A standard threshold is $p$=0.01, i.e. the false-alarm probability is 1\,\%. Based on the $\chi^2$, the $\chi^{2}_{\rm{red}}$ is calculated via

\begin{equation}
\label{chi_red}
\chi^2_{\rm{red}}(\vec{x})=\frac{\chi^2(\vec{x})}{d_f}
\end{equation}
where $d_f$ is the number of the degrees of freedom obtained from $d_f=N-P$ ($P$ number of parameters). In order to be considered a good fit, $\chi^2_{\rm{red}}$ should be of the order of unity.
 
In this work, we used two different approaches for comparing modeled spectra with the synthetic observations. Firstly, to fit the observations, we used the IDL fitting routine MPFIT \citep{markwardt2009}, which is an implementation of the nonlinear least-squares Levenberg-Marquardt (LM) algorithm \citep{more1978}. MPFIT uses the $\chi^2$ value to calculate best-fitting parameters. However, since it is based on a Newton-type method, it might find a local minimum which could be quite far from the actual global minimum of the $\chi^2$. Note that Newton-type solvers are to some extent dependent on the initial parameter guess.

Secondly, we calculate the  $\chi^2$ for a large number of parameter combinations to produce a  $\chi^2$ map of the parameter space. With these maps, local minima in the $\chi^2$ and potential degeneracies between model parameters could easily be visualized. Such a method was used, e.g., by \citet{lee2012}, \citet{cochran2011}, \citet{fabrycky2012}, or \citet{ford2012} in addition to formal fitting procedures. Error bars or uncertainties on retrieved parameters were then calculated from the $\chi^2$ maps.

For a large number of parameters (e.g., 8 in the full model described in Sect. \ref{theatmosphere}), producing $\chi^{2}$ maps is not practical, both in terms of calculation time and visualization. In such cases, the LM algorithm (or other  algorithms, such as Monte-Carlo) is essential to estimate parameter values.

\section{Results}

\label{resultsect}

\subsection{Surface conditions}

\label{sfccond}

In this section, we show the $\chi^2$ maps in the $T_{\rm{surf}}$-$p_{\rm{dry}}$ plane, i.e. the parameters relevant to surface habitability. The maps represent slices of the parameter space with all other 6 parameters (planetary mass, atmospheric composition, upper atmospheric T/p profile) held constant at the values stated in Table \ref{atmomodels}, i.e. the true parameters used when generating the synthetic spectra. This means that uncertainties on surface conditions cannot be assigned based on such slicing, since for real observations, the model parameters are not known exactly a priori. However, these $\chi^2$ maps illustrate the difficulty when trying to infer surface conditions from observations. The $\chi^{2}$ values were calculated for 20 synthetic observations, each with the specified S/N ratio per bin, but with a different realization of the noise. The median $\chi^{2}$ of these 20 observations is plotted to illustrate the general shape of the $\chi^2$ map at the chosen S/N ratio. Note that if we were to combine these 20 observations (equivalent to adding transits), we would increase the S/N ratio by a factor of $\sqrt{20}$.

\begin{figure}[h]
  \resizebox{\hsize}{!}{\includegraphics*{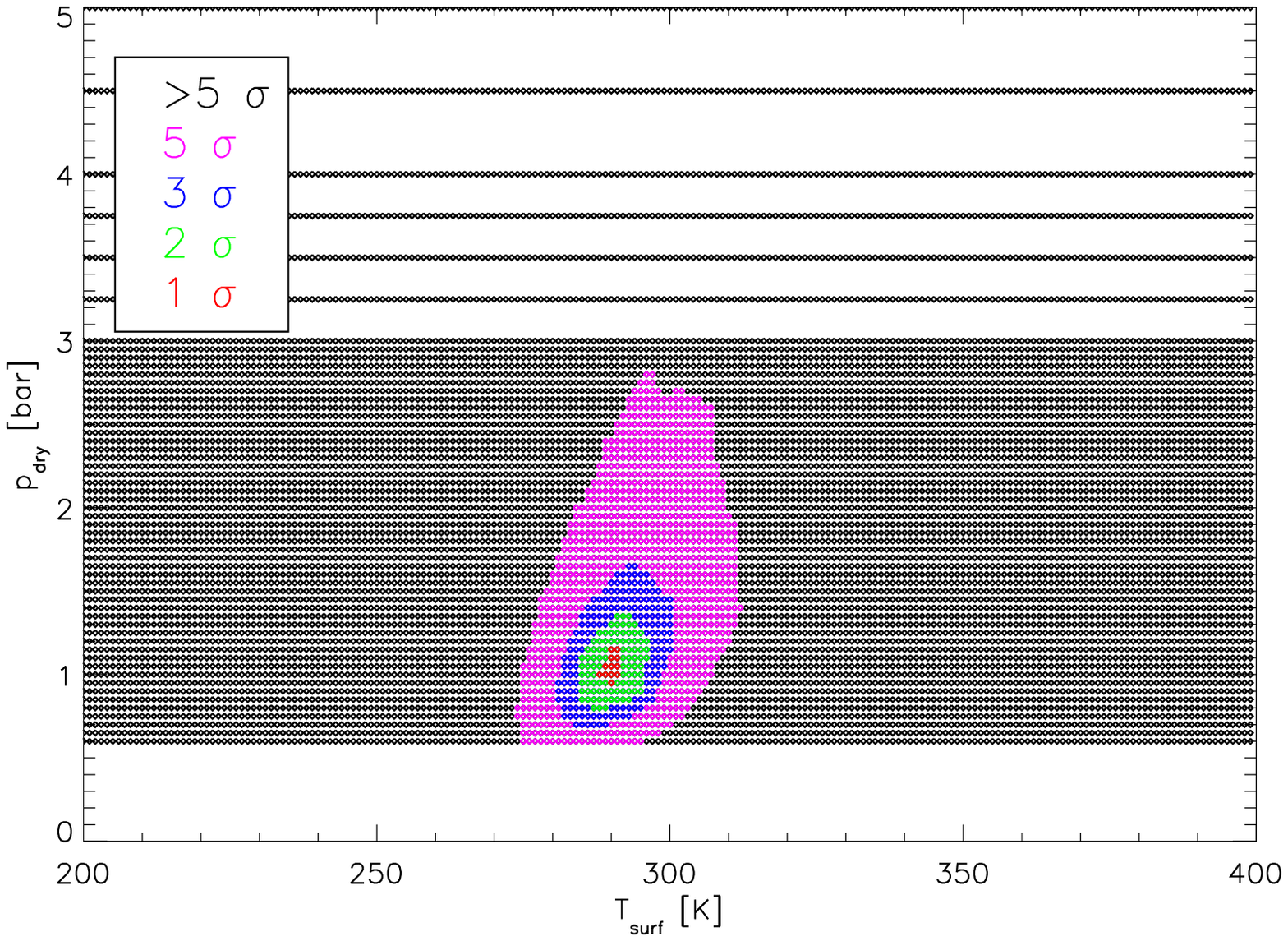}}
\resizebox{\hsize}{!}{    \includegraphics*{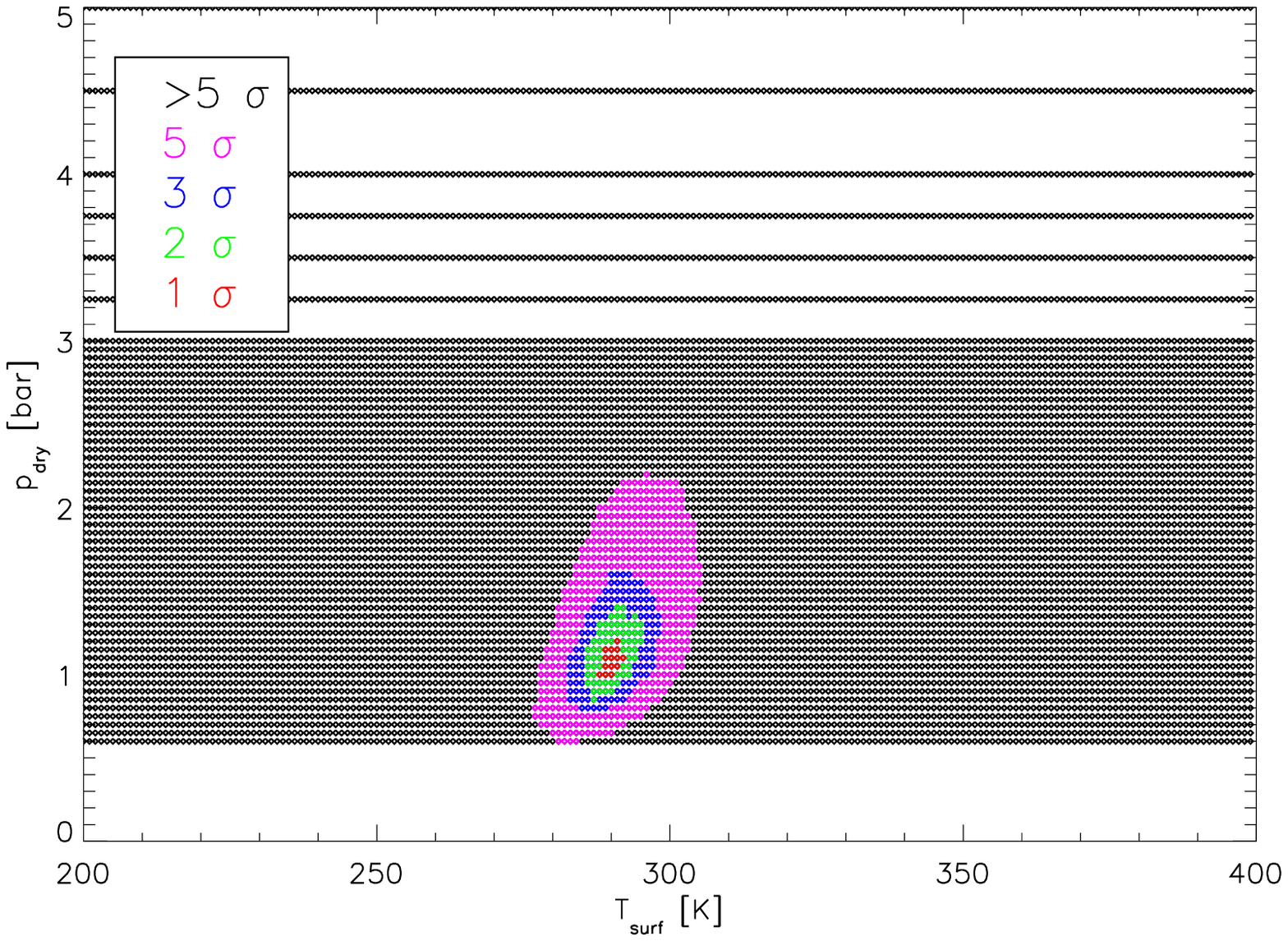}}\\
  \caption{$\chi ^2$ maps for EChO (\citealp{tinetti2012}, upper panel) and Darwin (\citealp{leger1996}, $R$=5) specifications. Symbols show considered grid points.}
  \label{surf_A1_map}
\end{figure}

Fig. \ref{surf_A1_map} shows the $\chi^2$ maps for the EChO (upper) and Darwin (lower) low resolution configuration, i.e. a spectral resolution of $R$=10 \citep{tinetti2012} and $R$=5 \citep{leger1996}, respectively. The mean S/N ratio per bin is approximately 5-6 in the upper part and 12 in the lower part, i.e. close to the respective S/N goals stated in Table \ref{missions}. The different colors in the $\chi^2$ maps represent increasing confidence levels, i.e. 1, 2, 3 and 5\,$\sigma$ confidence levels, corresponding to a distance of 1, 4, 9 and 25 to the minimum $\chi^2_{\rm{min}}$ (see Sect. \ref{fitmodel}). For EChO, at a 3\,$\sigma$ level (i.e., blue area in the $\chi^2$ maps), surface temperatures can be inferred to be higher than 280\,K and lower than 300\,K, indicating a habitable surface with relatively high confidence. However, for the extraordinary claim of the detection of (inhabited) habitable planets, one would need rather the 5\,$\sigma$ uncertainty level. This barely excludes surface temperatures below the freezing point of water, i.e. 273\,K in the EChO case, while surface habitability can be inferred at the 5\,$\sigma$ level for the presented Darwin specifications. In contrast, surface pressure is less constrained in both cases (about a factor of 4), but still within reasonable values for habitability.

\begin{figure}[h]
  \resizebox{\hsize}{!}{\includegraphics*{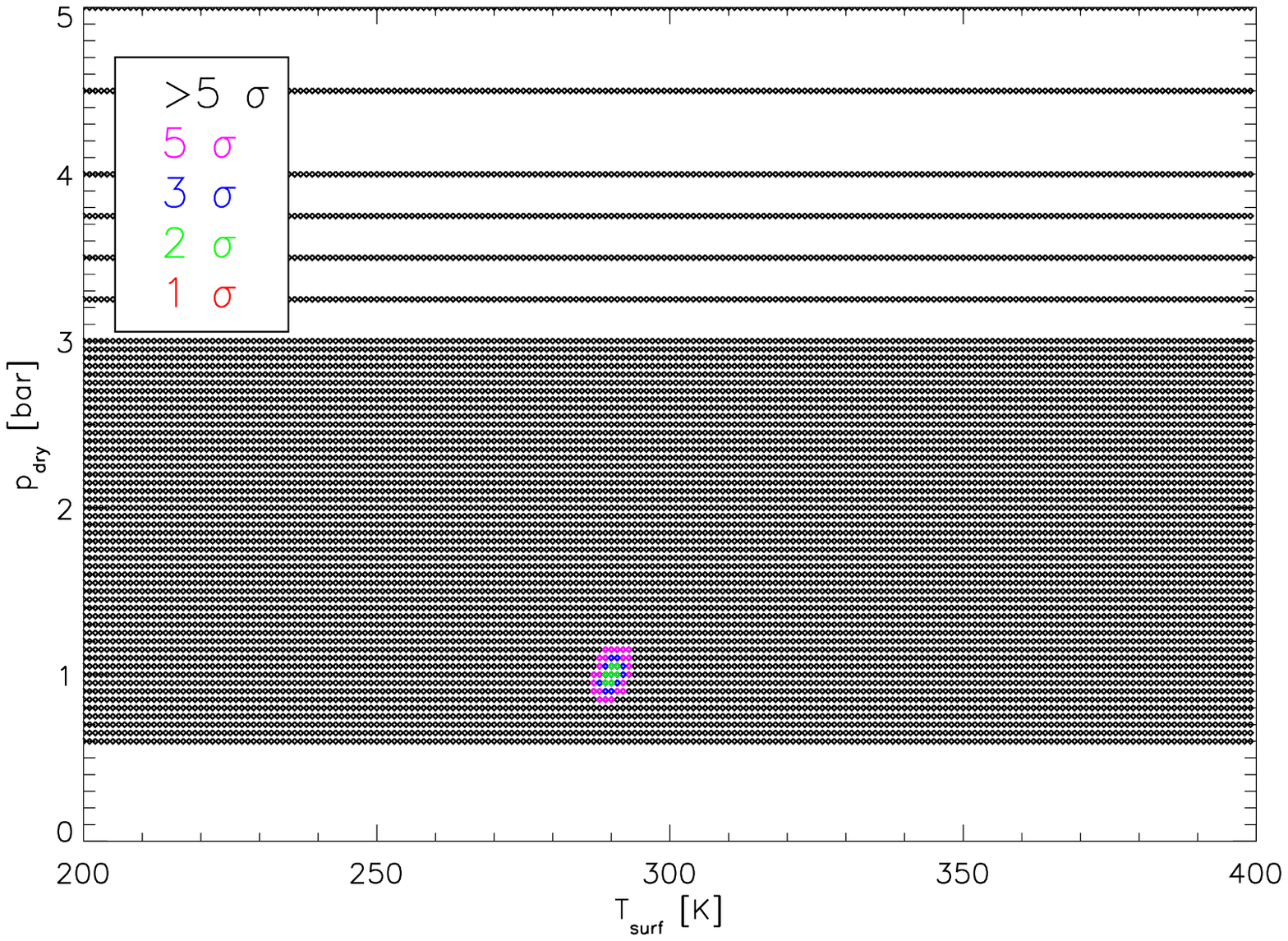}}\\
\resizebox{\hsize}{!}{    \includegraphics*{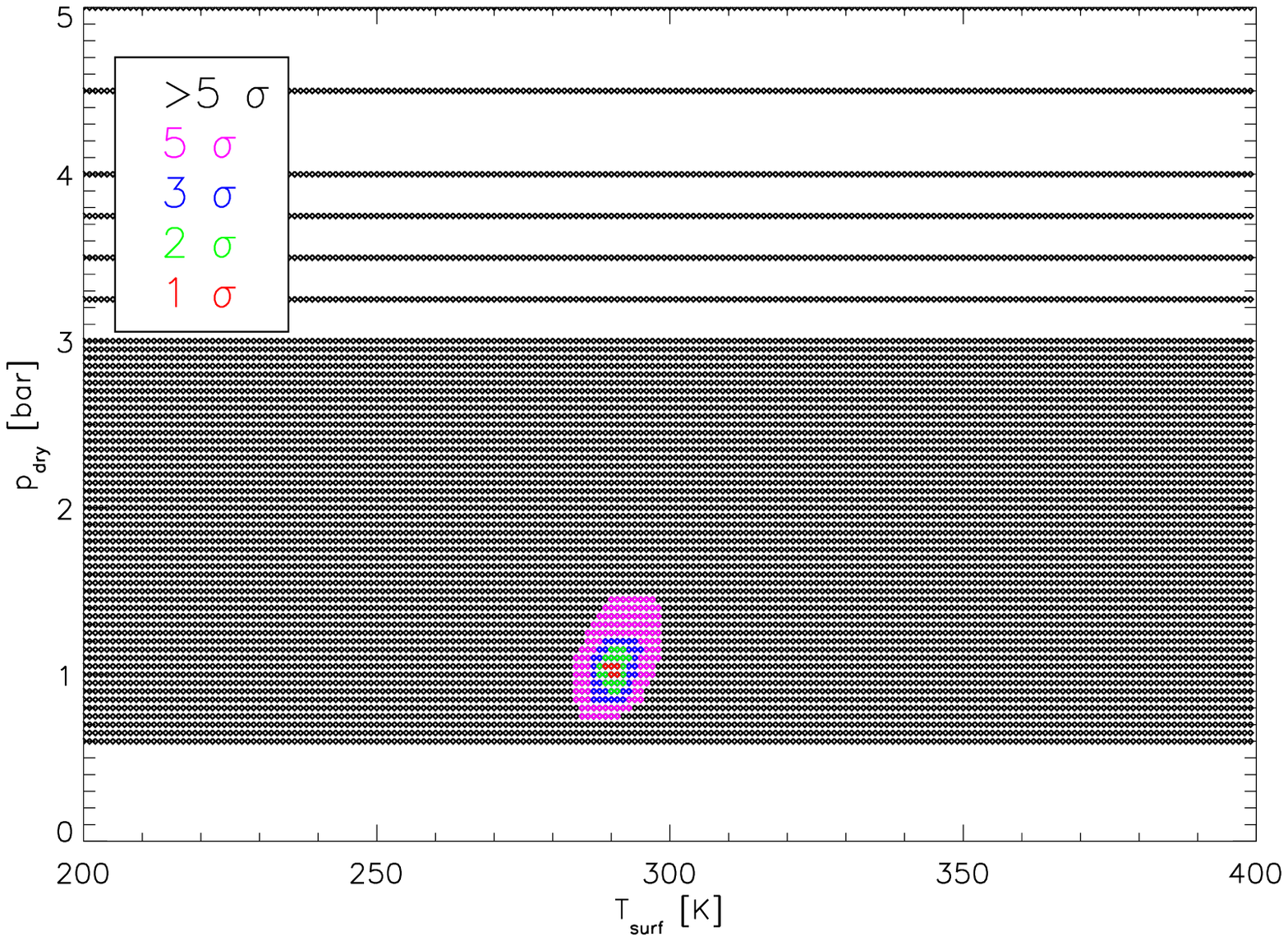}}\\
  \caption{$\chi ^2$ maps for Darwin spec ($R$=20, upper: \citealp{leger1996}, lower: \citealp{cockell2009exp}). Symbols show considered grid points.}
  \label{surf_A1_map_2}
\end{figure}

Fig. \ref{surf_A1_map_2} shows the $\chi^2$ maps for the medium resolution Darwin cases ($R$=20, \citealp{leger1996} and \citealp{cockell2009exp}). The mean S/N ratio per bin in these cases is 20 and 10, respectively (see stated S/N goals in Table \ref{missions}). It is clearly seen that the relatively high spectral resolution together with the high projected S/N ratio per bin allows for a secure $>$5\,$\sigma$ characterization of surface conditions (but keep in mind the idealized setup of the $\chi^2$ maps in this section).

In summary, in the idealized case presented here, the retrieval of surface conditions is possible with a high S/N ratio (of the order of 5 and more). The 1\,$\sigma$ uncertainties are relatively small, but the optimistic case considered here underestimates retrieval uncertainties. Adding free parameters such as atmospheric composition, stratospheric temperature structure and planetary mass to the fit will only increase the uncertainties of the retrieval. Furthermore, it should be kept in mind that the cloud-free nature of our model implies that retrieval possibilities for surface conditions might be over-estimated.

\subsection{Atmospheric composition}

\label{atmoret}

The detection of atmospheric species is another major goal for exoplanet characterization. Habitability, in terms of surface conditions, does not necessarily imply an inhabited planet, hence the search for atmospheric biomarker species holds the greatest potential for actually finding extraterrestrial life. Therefore, in this section, we show the $\chi^2$ maps for the atmospheric composition (i.e., $h$, CO$_{2}$ and O$_{3}$), again, as above, with all other parameters held constant at the true values. As before in Sect. \ref{sfccond}, the different colors in the $\chi^2$ maps represent increasing confidence levels and the median $\chi^{2}$ of the 20 synthetic observations is plotted.

\begin{figure}[h]
  \resizebox{\hsize}{!}{\includegraphics*{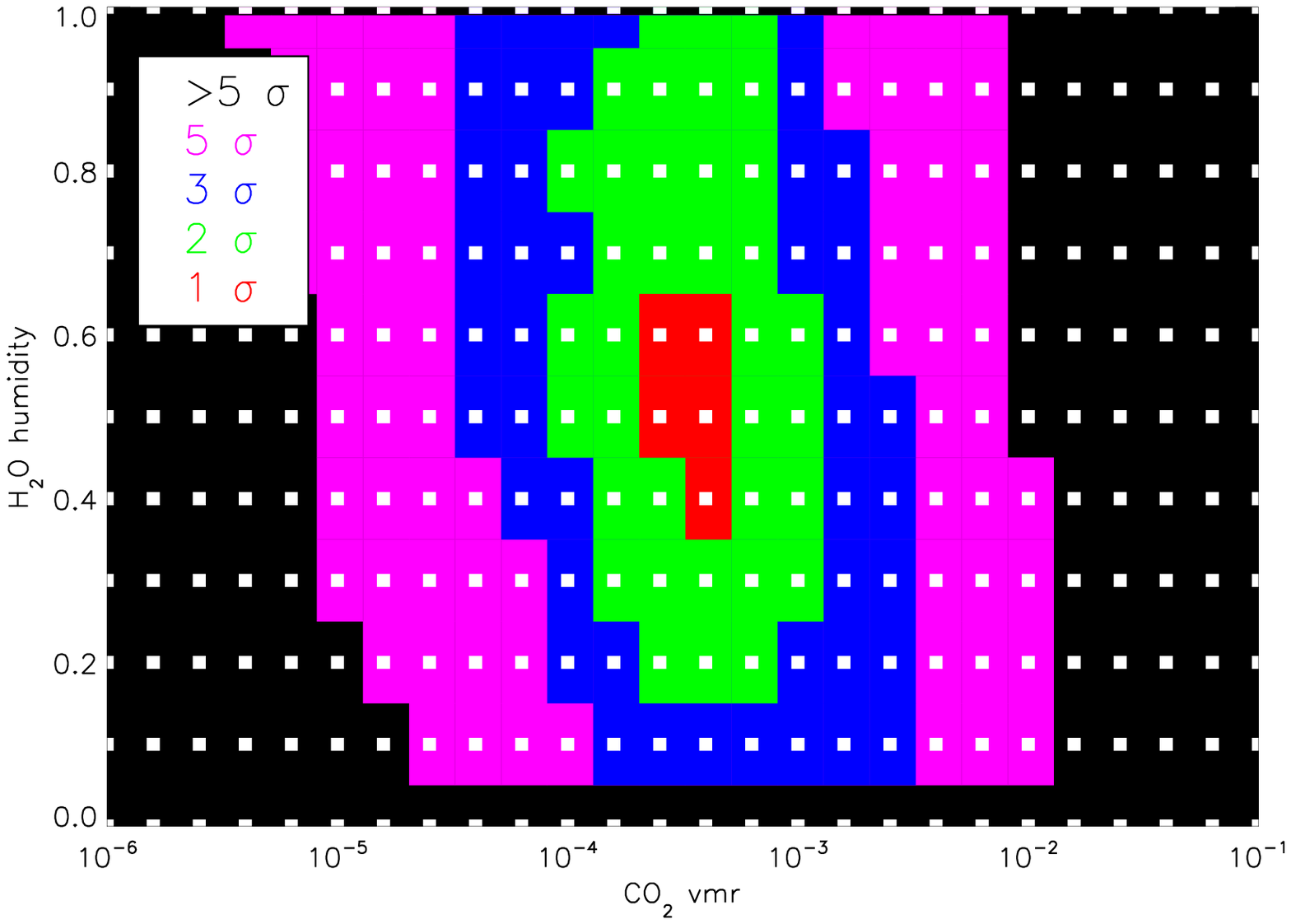}}\\
\resizebox{\hsize}{!}{\includegraphics*{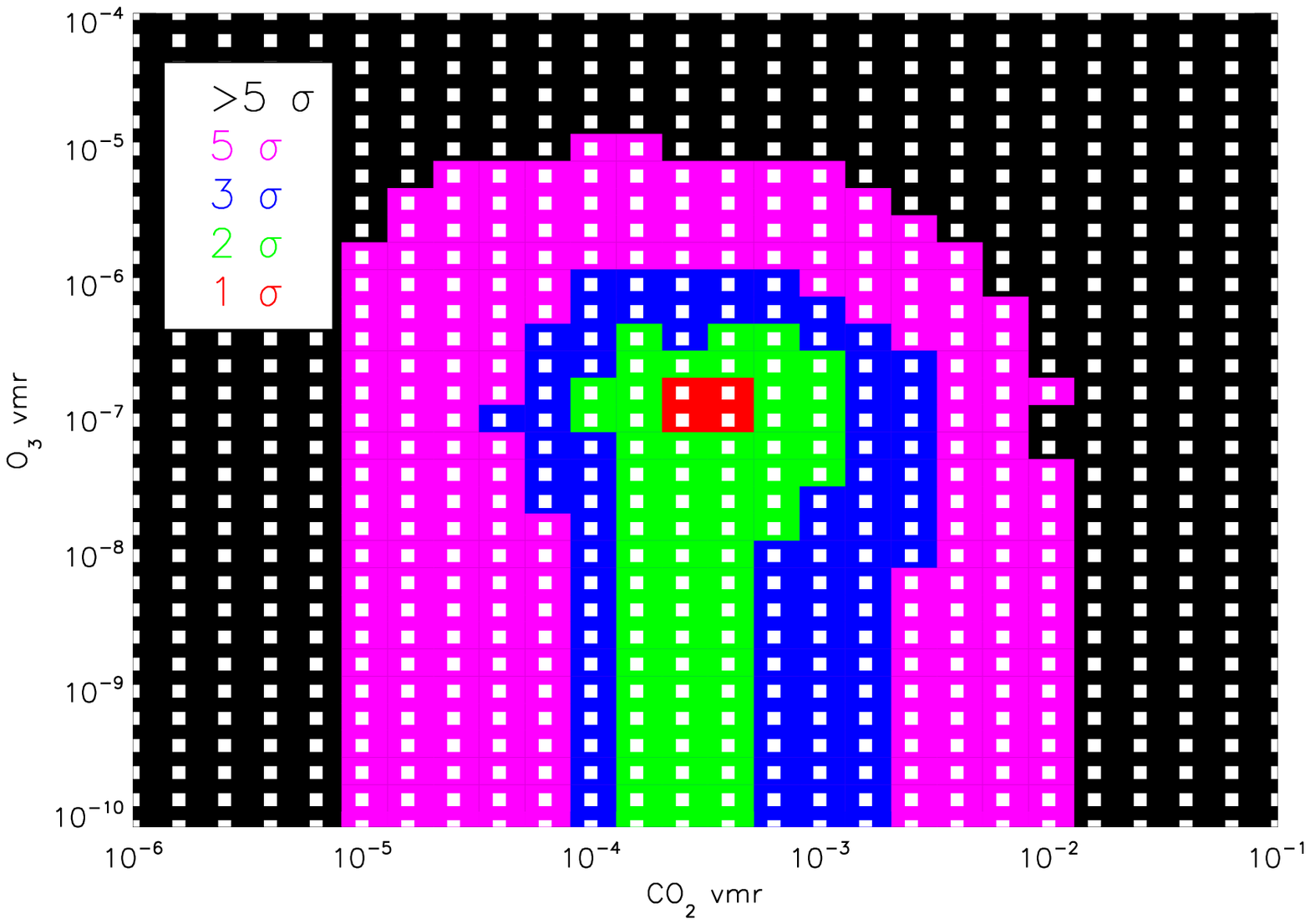}}\\
\caption{$\chi^{2}$ contours for ECho specifications from Table \ref{missions}. White symbols show considered grid points.}
  \label{comp_chi2_A1_spec}
\end{figure}

Figure \ref{comp_chi2_A1_spec} shows the $\chi^{2}$ maps for the EChO resolution ($R$=10) in the respective CO$_{2}$-$h$ (upper) and CO$_{2}$-O$_{3}$ (lower) plane. The S/N ratio is chosen to be equal to the EChO goal of 5 (see Table \ref{missions}). It is clearly seen that the presence of water can be securely inferred from the spectrum, as $h$=0 is excluded at high confidence. Also, CO$_2$ can be detected clearly, since the minimum CO$_2$ mixing ratio is about 10$^{-5}$. O$_3$ is detected at the 1\,$\sigma$ level, but O$_3$ concentrations of 10$^{-10}$ are compatible with the observations at the 2\,$\sigma$ level, hence the detection would be somewhat marginal. Note however that even 2\,$\sigma$ constraints would be very useful for the potential selection of candidates which would be subjected to further, more detailed follow-up observations.

However, the goal of exoplanet space missions (such as EChO, Darwin, Finesse) is not only the detection of atmospheric species (i.e., strong lower limits), but also the characterization (i.e. a quantitative estimate of the concentration, meaning strong upper limits). As implied by Fig. \ref{comp_chi2_A1_spec}, 5\,$\sigma$ uncertainties on CO$_{2}$ concentration cover about three orders of magnitude, and the humidity is only constrained to about a factor of 9 (with $h$=1 and $h$=0 excluded at high confidence). Therefore, it seems unlikely that a S/N ratio of about 5 at spectral resolution $R$=10 will allow for an accurate characterization of Earth-like atmospheres.

\begin{figure}[h]
 \resizebox{\hsize}{!}{  \includegraphics*{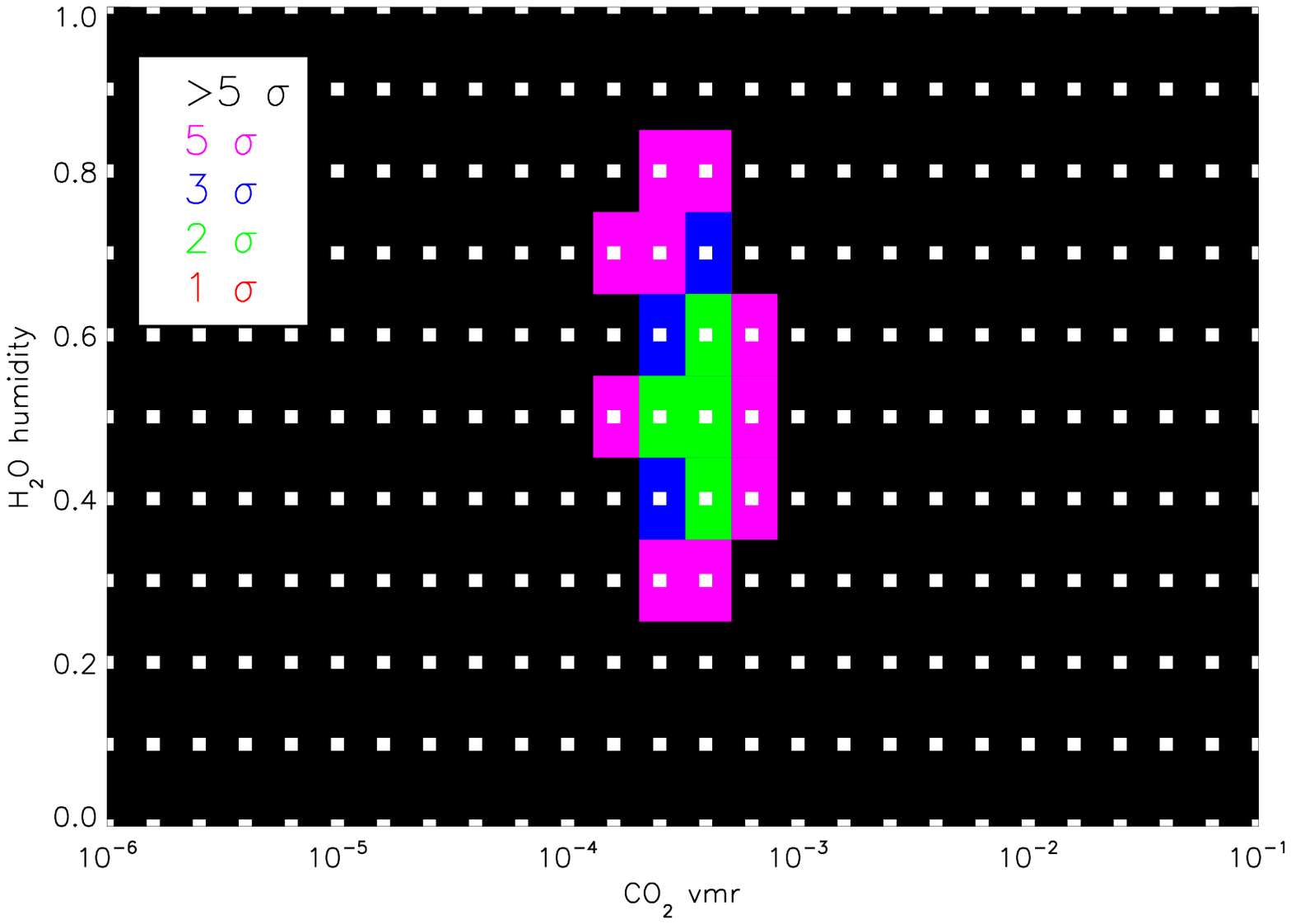}}\\
\resizebox{\hsize}{!}{  \includegraphics*{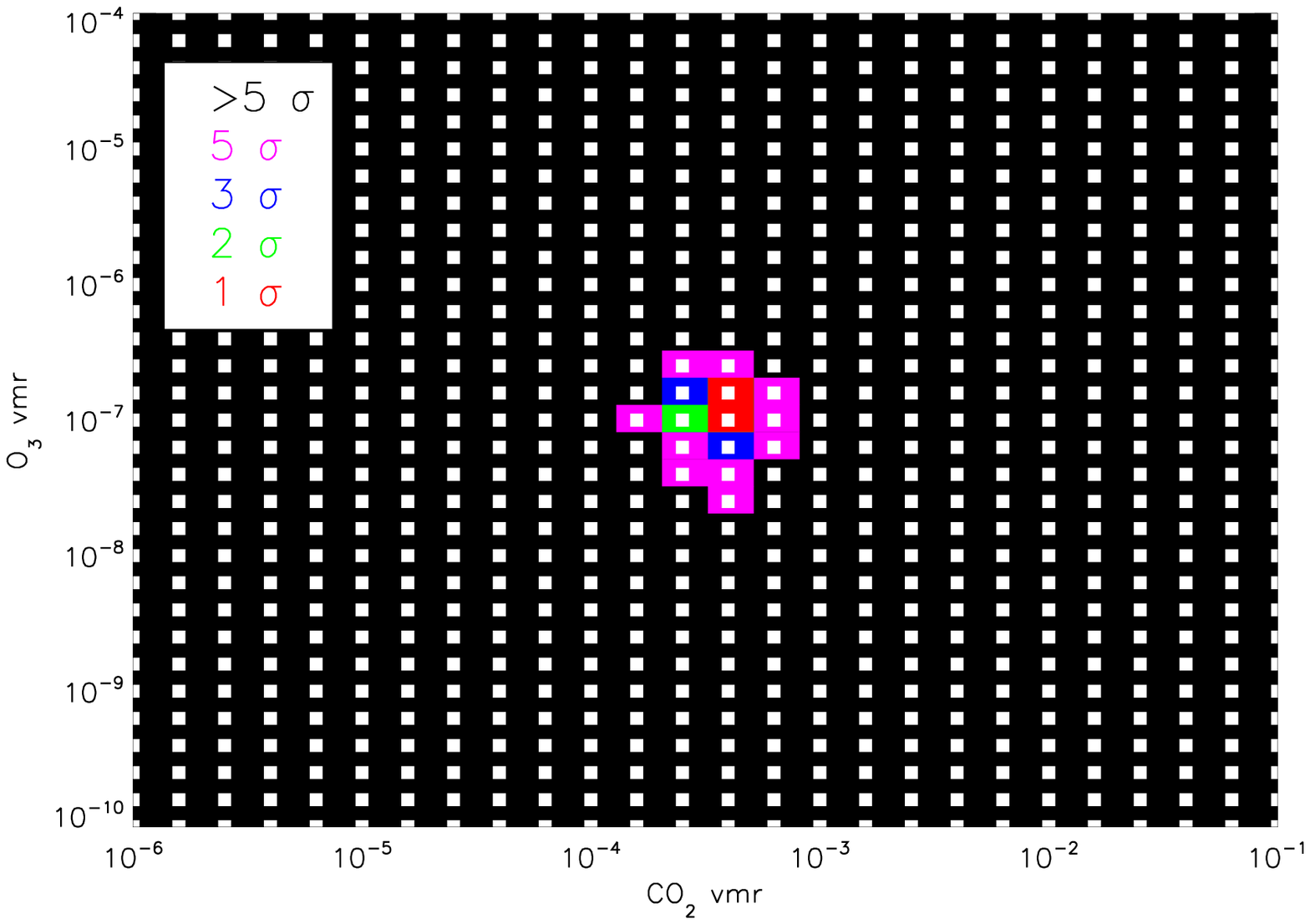}}\\
\caption{$\chi^{2}$ contours for ECho resolution, with increased S/N ratio of 28. White symbols show considered grid points.}
  \label{comp_chi2_A1_need}
\end{figure}

In Fig. \ref{comp_chi2_A1_need}, we show the $\chi^2$ maps with the same spectral resolution ($R$=10) but with a S/N ratio sufficient to limit the 5\,$\sigma$ uncertainties of the atmospheric composition to below one order of magnitude. The S/N ratio per bin ($\sim$28) is five times higher than the value stated in Table \ref{missions}, but allows for the characterization of H$_2$O, CO$_2$ and O$_3$ in this ideal model exercise (all parameters except composition fixed). An increase of S/N ratio from 5 to 28 translates into a factor of about 30 in integration time if read-out noise is not the dominant noise source, because then, S/N scales with the square root of the integration time. 

\begin{figure}[h]
\resizebox{\hsize}{!}{  \includegraphics*{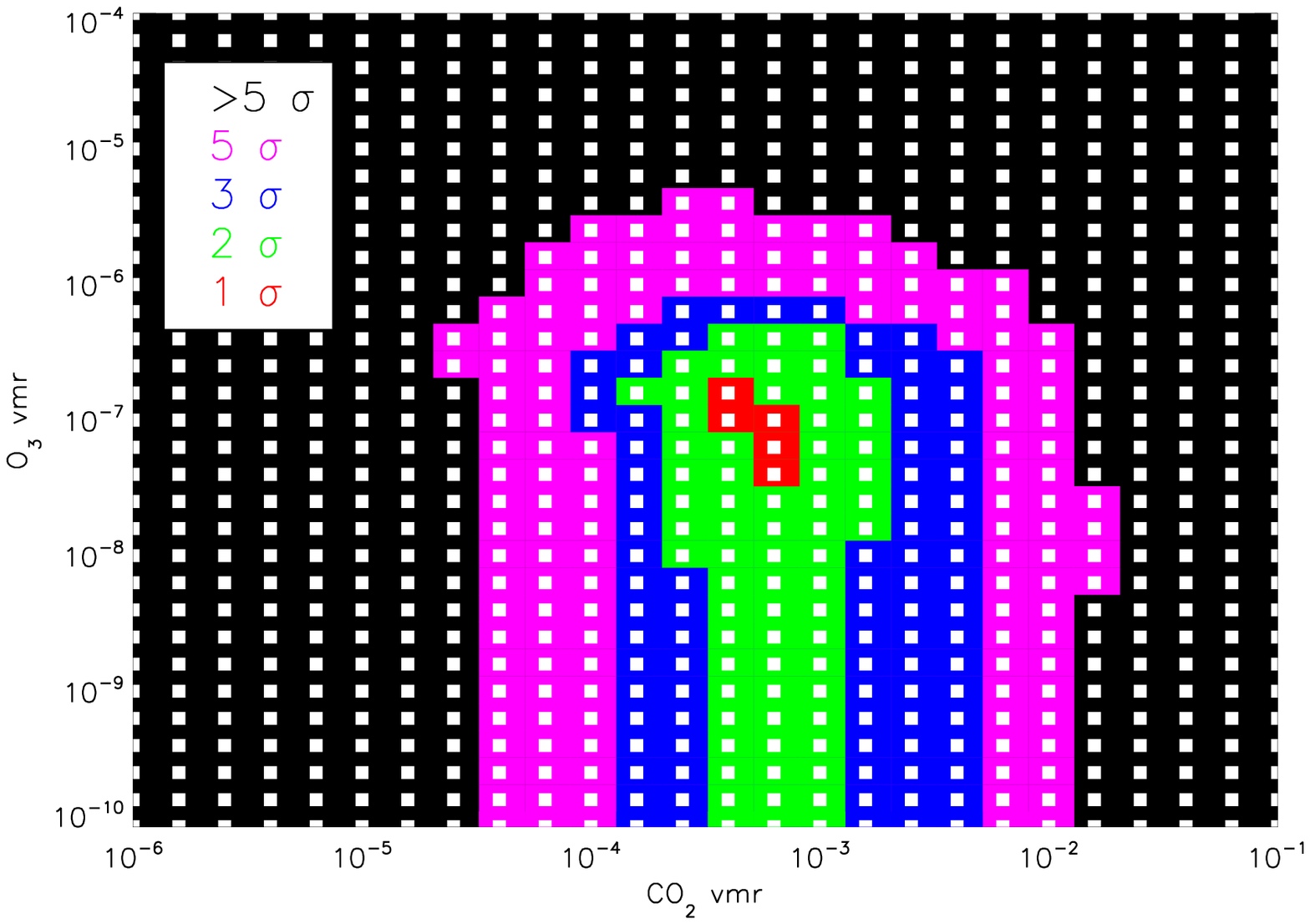}}\\
\resizebox{\hsize}{!}{  \includegraphics*{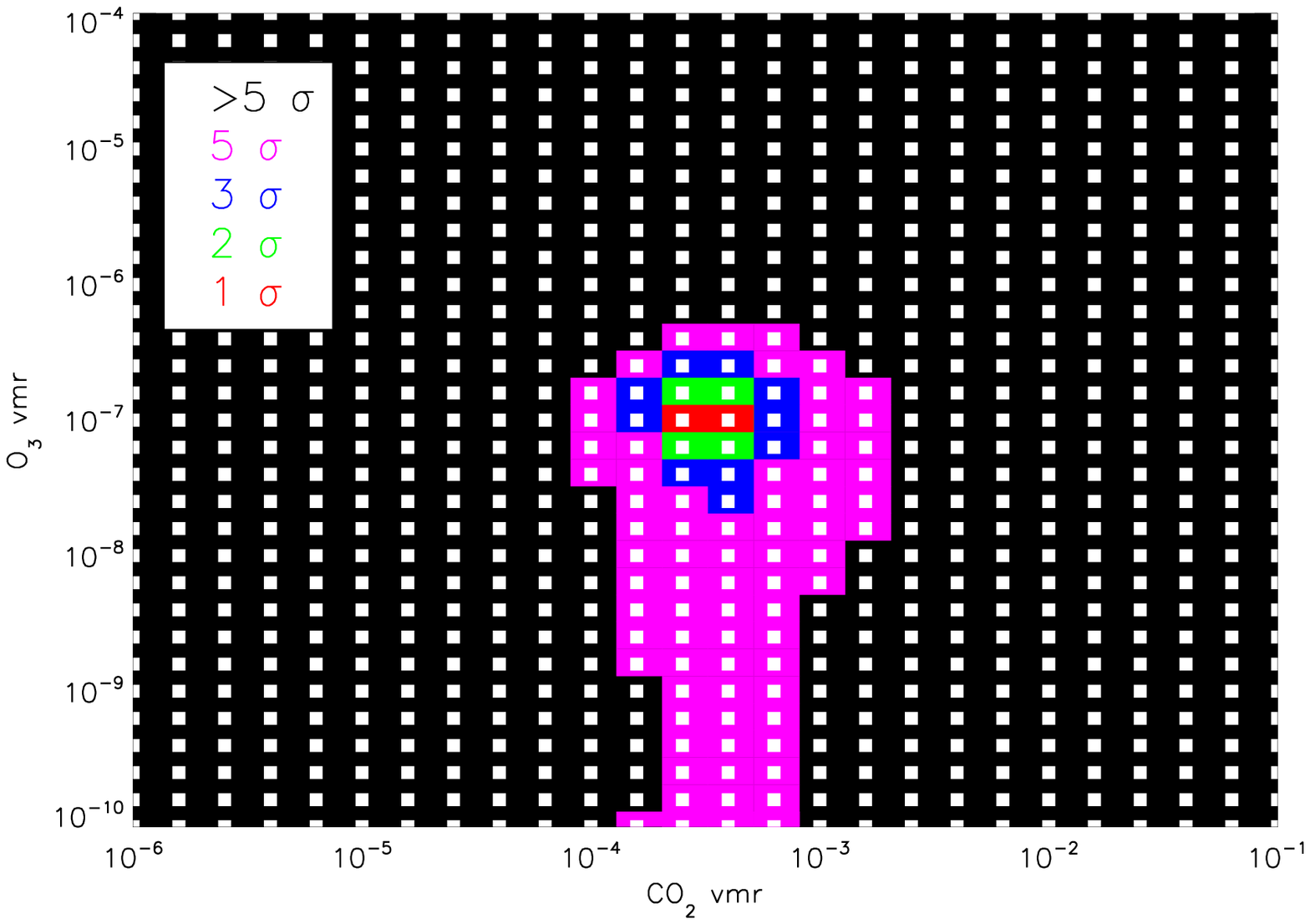}}\\
\caption{$\chi^{2}$ contours for Darwin low resolution specification (upper) and needed S/N ratio (lower). White symbols show considered grid points.}
  \label{comp_chi2_A1_2}
\end{figure}

\begin{figure}[h]
 \resizebox{\hsize}{!}{ \includegraphics*{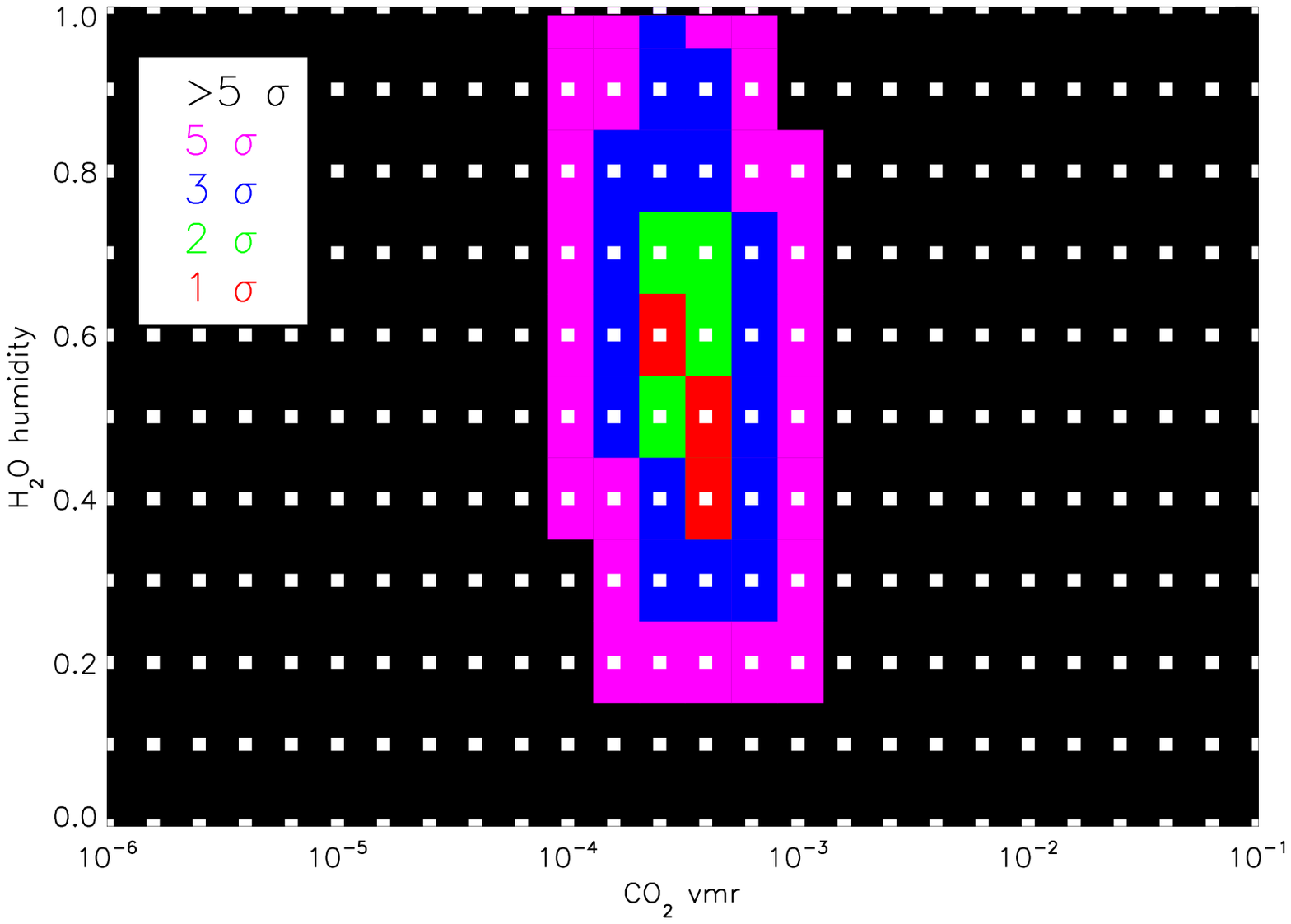}
  \includegraphics*{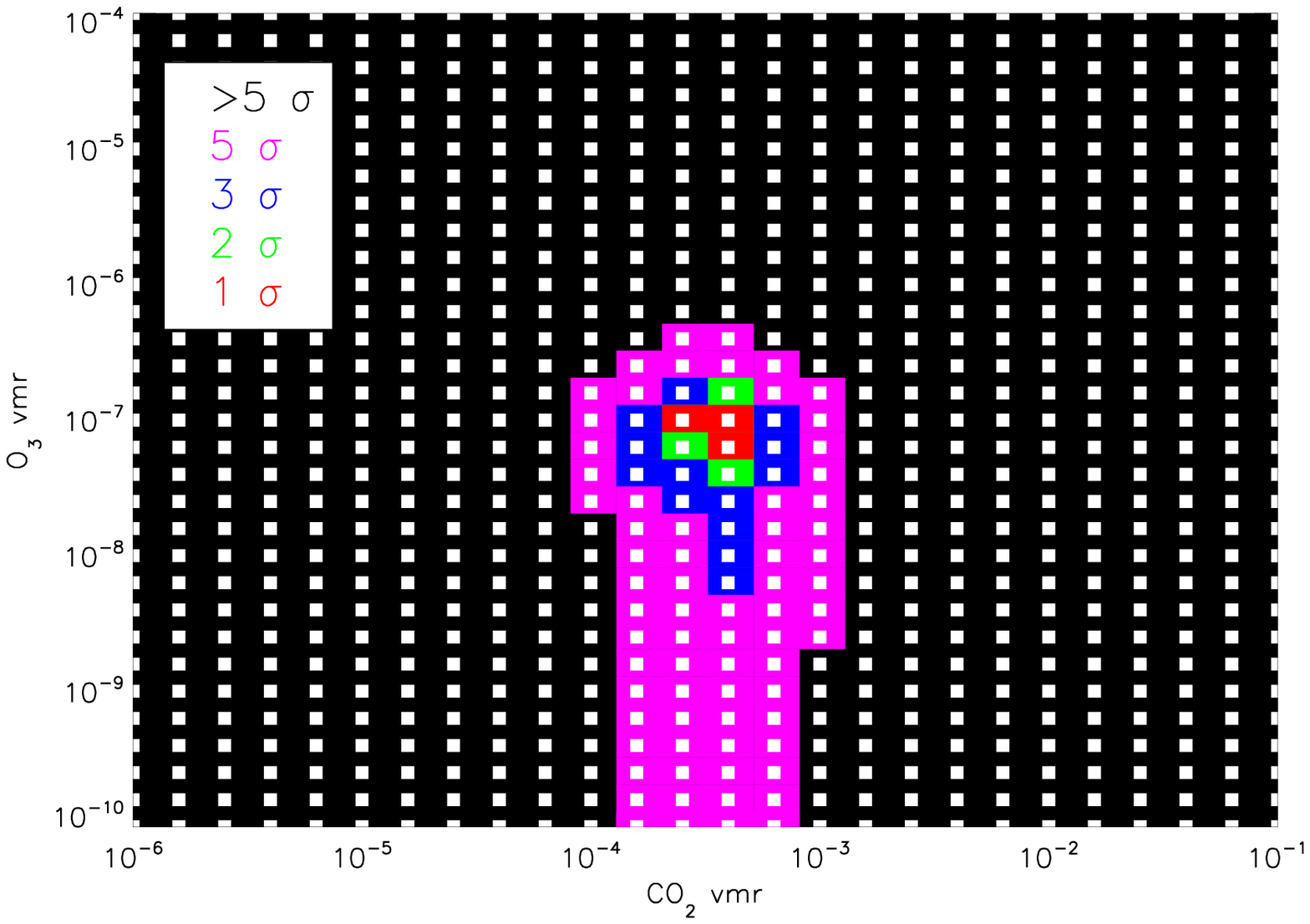}}\\ 
\resizebox{\hsize}{!}{  \includegraphics*{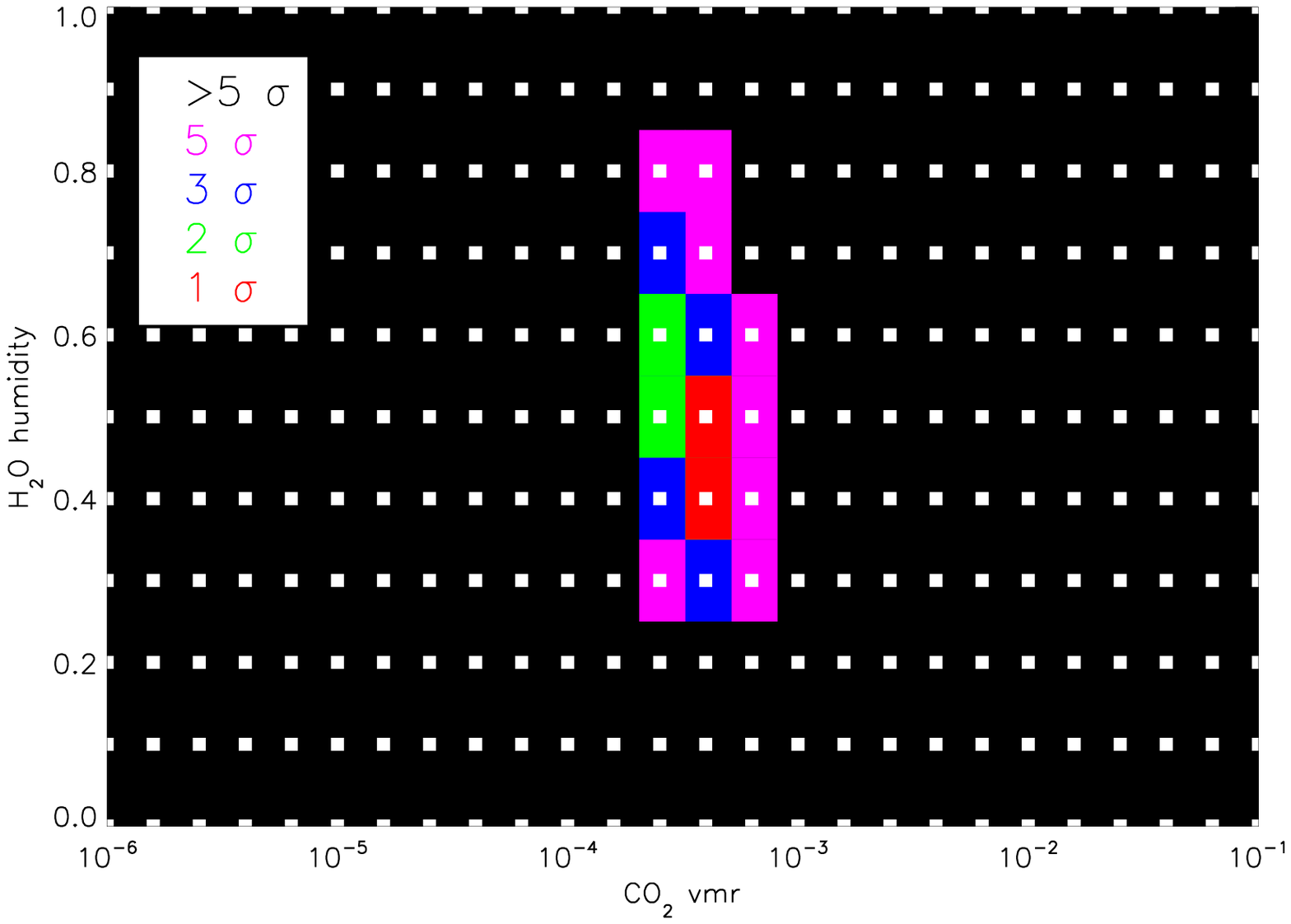}
  \includegraphics*{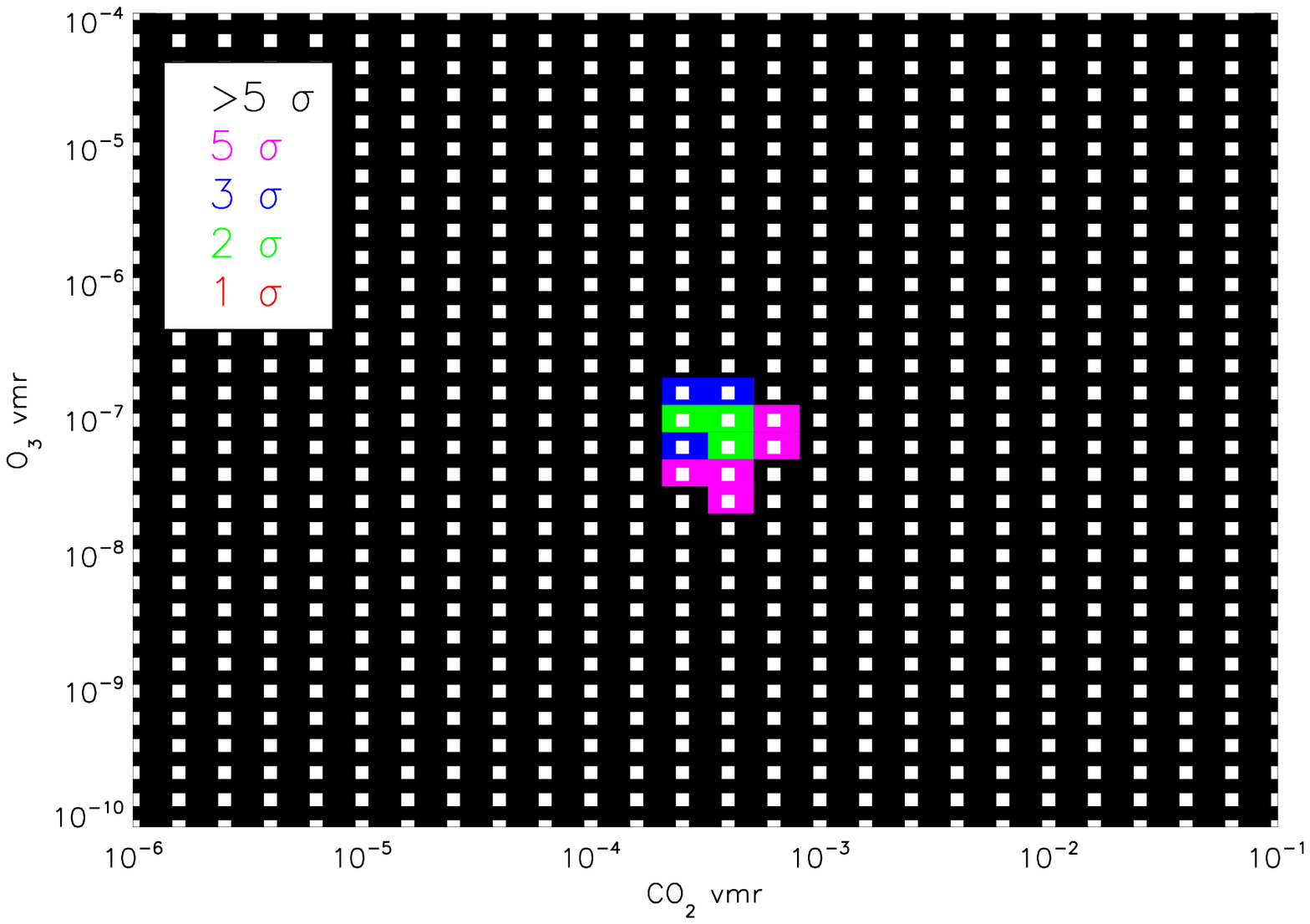}}\\
\caption{$\chi^{2}$ contours for Darwin high resolution specifications (upper) and needed S/N ratios (lower). White symbols show considered grid points.}
  \label{comp_chi2_A1_3}
\end{figure}

In Figs. \ref{comp_chi2_A1_2} and \ref{comp_chi2_A1_3}, we show analogous $\chi^2$ maps for the Darwin low-resolution ($R$=5, see Table \ref{missions}) and medium-resolution ($R$=20, see Table \ref{missions}) case. Since at low resolution, the aim is to constrain CO$_2$, we only show the CO$_2$-O$_3$ plane in Fig \ref{comp_chi2_A1_2}. Results in Fig. \ref{comp_chi2_A1_2} suggest that a S/N ratio of around 12 is not sufficent at $R$=5 to constrain CO$_2$ better than about 2 orders of magnitude. To obtain one order of magnitude uncertainty at the 5\,$\sigma$ level, an increase in S/N ratio to more than 25 is needed (lower part of Fig. \ref{comp_chi2_A1_2}).

At medium spectral resolution of $R$=20, \citet{cockell2009exp} and \citet{cockell2009asbio} state that an S/N of 10 is aimed for to characterize the atmosphere with respect to CO$_2$, H$_2$O and O$_3$. Hence, we show both the CO$_2$-$h$ (left column) and the CO$_2$-O$_3$ (right column) plane in Fig. \ref{comp_chi2_A1_3}. As can be inferred from the upper panel (S/N ratio of 10) in Fig. \ref{comp_chi2_A1_3}, except for CO$_2$, atmospheric composition cannot be constrained with these specifications. An increase to a S/N ratio of about 20 (lower panel) is needed to better constrain H$_2$O and O$_3$ at the 5\,$\sigma$ level (although O$_3$ is already constrained at the 3\,$\sigma$ level for an S/N ratio of 10). Note that this S/N ratio at $R$=20 corresponds to the stated O$_3$ characterization aim of \citet{leger1996} in Table \ref{missions}.

An increase of about a factor of 2 in S/N ratio then translates into a factor 4 for Darwin in terms of exposure time necessary for a specific target. This significantly reduces the number of potential target stars.

\begin{figure}[h]
 \resizebox{\hsize}{!}{ \includegraphics*{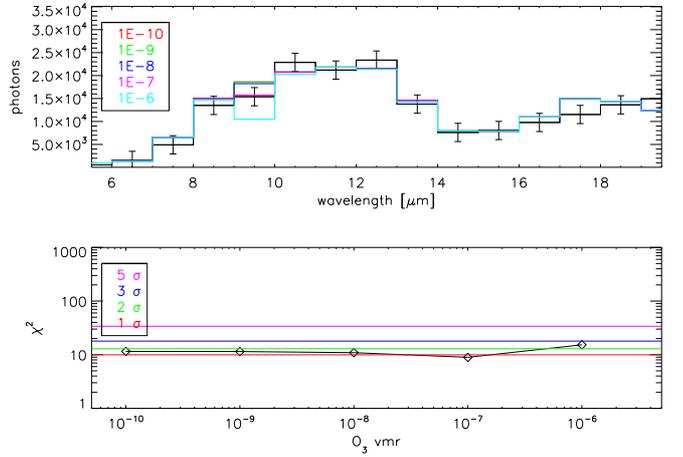}}\\
\caption{Influence of O$_3$ on synthetic observations (upper panel, noisy spectrum in black) and corresponding $\chi^2$ values (lower panel, diamonds). EChO resolution and S/N ratio.}
  \label{o3_1}
\end{figure}

\begin{figure}[h]
\resizebox{\hsize}{!}{  \includegraphics*{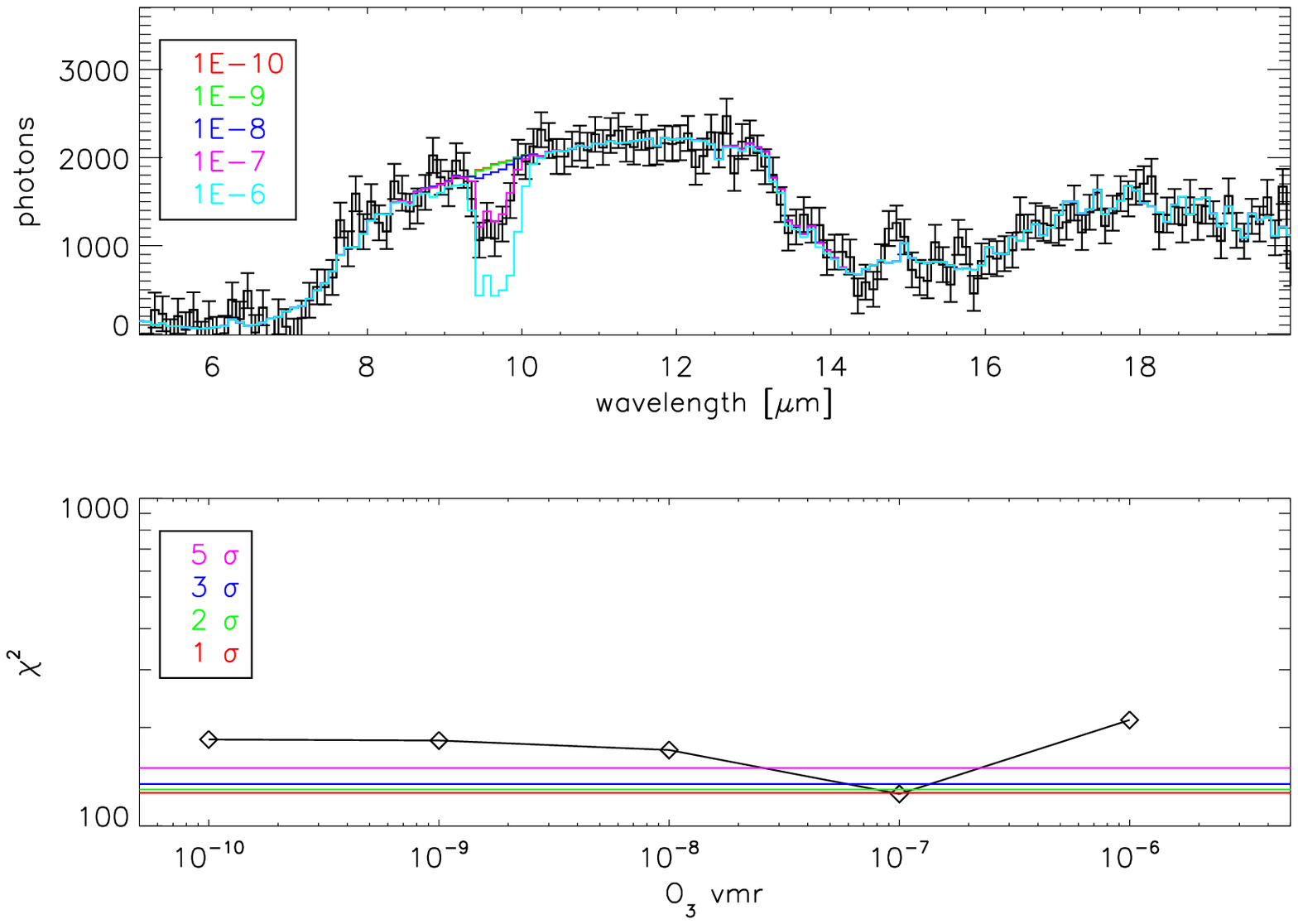}}\\
\caption{Influence of O$_3$ on synthetic observations (upper panel, noisy spectrum in black) and corresponding $\chi^2$ values (lower panel, diamonds). $R$=100 and S/N ratio of 6.}
  \label{o3_2}
\end{figure}

To further illustrate the difficulty to obtain firm constraints on O$_3$ concentration, we show in Figs. \ref{o3_1} and \ref{o3_2} sample noisy observations together with spectra of model atmospheres containing various amounts of O$_3$, but otherwise the 7 remaining parameters fixed at the ``correct'' values. Fig. \ref{o3_1} shows a synthetic observation at $R$=10 and a mean S/N ratio per bin of about 6, i.e. the current EChO specifications. It is clearly seen that increasing the O$_3$ concentration from 10$^{-10}$ to 10$^{-6}$ changes the O$_3$ band (within 2\,$\sigma$ in the bin), however, the change in overall $\chi^2$ is not significant, as expected from Fig \ref{comp_chi2_A1_spec}.

In Fig. \ref{o3_2}, the effect of O$_3$ is shown with the same S/N ratio of about 6, but at a spectral resolution of $R$=100. As expected, the higher resolution clearly improves upon the accuracy for O$_3$ retrieval. 

Therefore, our results imply that the current Darwin design is probably overly optimistic with respect to the number of habitable terrestrial planets which could be characterized and searched for atmospheric biomarkers with emission spectroscopy. Even putting aside the current paucity of targets, since no transiting HZ planet are known so far and only a few candidate super-Earths in or near the HZ have been discovered (see Introduction), EChO is probably also unlikely to constrain atmospheric composition with the current S/N and $R$ goals.

\begin{table*}
  \caption{Found best-fit parameters after 100 randomly initialized fits, with associated 3\,$\sigma$ uncertainties. Spectral resolution and S/N ratio as proposed for EChO and Darwin, as well as increased S/N ratios. Parameters which could be fitted reasonably in red. Statistics of the fit results are also included (see Sect. \ref{fitmodel}).}\label{echofound}
\begin{tabular}{l|c|c||c|c||c|c}
\hline
\hline
parameter&  $R$=10, S/N=5& $R$=10, S/N=28 & $R$=5, S/N=12 &$R$=5, S/N=26& $R$=20, S/N=10& $R$=20, S/N=20\\
\hline 
\hline 
$T_{\rm{TOA}}$ [K]     & 344 $^{+6}_{-144}$ &  300 $^{+50}_{-100}$ & 289 $^{+61}_{-89}$ & 200 $^{+150}_{-0}$& 349 $^{+1}_{-149}$& 309 $^{+41}_{-66}$\\
\hline 
{\color{red}$T_{\rm{surf}}$} [K]    & 285 $^{+115}_{-19}$ &  289 $^{+8}_{-8}$ & 276 $^{+123}_{-19}$&297 $^{+37}_{-13}$&285 $^{+13}_{-12}$&295 $^{+7}_{-7}$\\
\hline 
$p_{\rm{trop}}$ [bar]    & 0.75 $^{+0.0}_{-0.7}$ & 0.75 $^{+0.0}_{-0.67}$ & 0.75 $^{+0.0}_{-0.74}$& 0.49 $^{+0.26}_{-0.48}$& 0.17 $^{+0.58}_{-0.10}$& 0.75 $^{+0}_{-0.67}$\\
\hline 
{\color{red}$p_{\rm{dry}}$} [bar]    &4.5 $^{+9.1}_{-4}$ &  2.7 $^{+1.2}_{-2.2}$ & 2.9 $^{+22.1}_{-2.4}$& 1.7 $^{+23.3}_{-1.2}$& 1.1 $^{+4.2}_{-0.6}$& 2.9 $^{+0.9}_{-2.4}$\\
\hline 
$h$    & 1 $^{+0}_{-0.9}$ & 0.23 $^{+0.77}_{-0.15}$& 1 $^{+0}_{-0.99}$&0.15 $^{+0.85}_{-0.14}$&1 $^{+0}_{-0.95}$&0.10 $^{+0.82}_{-0.08}$\\
\hline 
{\color{red}CO$_2$ vmr}    & 6.4$\cdot10^{-5}$ $^{+0.98}_{-5.9\cdot10^{-5}}$&  5.7$\cdot10^{-5}$ $^{+1.4\cdot10^{-4}}_{-3.3\cdot10^{-5}}$& 8.8$\cdot10^{-4}$ $^{+0.1}_{-8.8\cdot10^{-4}}$&4.9$\cdot10^{-5}$ $^{+5.7\cdot10^{-3}}_{-4.8\cdot10^{-5}}$&4.0$\cdot10^{-4}$ $^{+2.9\cdot10^{-3}}_{-3.8\cdot10^{-4}}$&4.1$\cdot10^{-5}$ $^{+1.0\cdot10^{-3}}_{-2.0\cdot10^{-5}}$\\
\hline 
O$_3$ vmr     &3.0$\cdot10^{-8}$ $^{+6.0\cdot10^{-6}}_{-3\cdot10^{-8}}$&  4.3$\cdot10^{-8}$ $^{+4.4\cdot10^{-7}}_{-1.6\cdot10^{-8}}$& 1.4$\cdot10^{-8}$ $^{+2.8\cdot10^{-3}}_{-1.4\cdot10^{-8}}$& 7.6$\cdot10^{-8}$ $^{+2.2\cdot10^{-6}}_{-7.3\cdot10^{-8}}$&1.1$\cdot10^{-7}$ $^{+3.6\cdot10^{-7}}_{-9.9\cdot10^{-8}}$&2.9$\cdot10^{-8}$ $^{+3.3\cdot10^{-7}}_{-1.0\cdot10^{-8}}$\\
\hline 
{\color{red}$m_p$ }[$M_{\oplus}$]    & 1.28 $^{+0.51}_{-0.72}$ &  0.97 $^{+0.28}_{-0.21}$ & 1.47 $^{+0.91}_{-0.94}$&0.81 $^{+0.41}_{-0.21}$&1.17 $^{+0.38}_{-0.44}$&0.82 $^{+0.23}_{-0.15}$\\
 \hline
 $N$    & 15  &  15 & 8  & 8  & 30 & 30\\
 \hline
 $d_f$  &7 & 7 & 0 & 0 & 22 & 22\\
\hline
$\chi^2_{\rm{min}}$ & 6.33 & 3.99  & 1.17 & 3.90 & 24.81 & 16.77\\
 \hline
$\chi^2_{\rm{red}}$ & 0.90 & 0.57 & $\infty$& $\infty$& 1.12& 0.76\\
\hline
$\chi^2_p$ for $p$=0.01& 30.57 & 30.57 & 20.09 & 20.09 & 50.89 & 50.89\\
\hline
\% of fits  $\chi^2$$<$$\chi^2_p$& 86 & 57 & 76 & 57 & 70 & 70\\
 \hline
\end{tabular}
\end{table*}

\subsection{Complete retrieval of planetary parameters}

\label{noconstraints}

In this section, we performed a non-linear least squares fit to one specific synthetic observation using MPFIT (see above, Sect. \ref{fitmodel}). All 8 model parameters were allowed to vary, as is the case for real observations, and we initialized the fit with a random guess of the parameters. Upper and lower boundaries for the fit parameters are stated in Table \ref{bound}. The numerical values for the parameter boundaries are mostly motivated by the fact we aim at investigating habitable, terrestrial planets, which, e.g., puts constraints on the planetary mass and the surface temperature. In total, 100 randomly initialized fits were performed.

\begin{table}[h]
  \caption{Parameter range used for initial guess and retrieval iteration.}\label{bound}
\begin{tabular}{l|c}
\hline
\hline
parameter & range \\
\hline 
\hline 
$T_{\rm{TOA}}$ [K]     & 200 -- 350\\
$T_{\rm{surf}}$ [K]    & 200 -- 400\\
$p_{\rm{trop}}$ [bar]    & 0.01 -- 0.75 \\
$p_{\rm{dry}}$ [bar]    & 0.5 -- 25 \\
$h$    & 0 -- 1 \\
CO$_2$ vmr    & 0 -- 0.99 \\
O$_3$ vmr     & 0 -- 0.01 \\
$m_p$ [$M_{\oplus}$]    & 0.01 -- 10 \\
 \hline
\end{tabular}
\end{table}

Minimizing the $\chi^2$ determines the best-fitting, optimal parameter vector $\vec{x_{\rm{opt}}}$ corresponding to $\chi^2_{\rm{min}}$. The 100 fit results were used to determine $\vec{x_{\rm{opt}}}$. Uncertainties were then estimated based on the distance $\Delta \chi^2$ to $\chi^2_{\rm{min}}$ (see Sect. \ref{fitmodel}). 

\begin{figure}[h]
1  
 \resizebox{\hsize}{!}{ \includegraphics{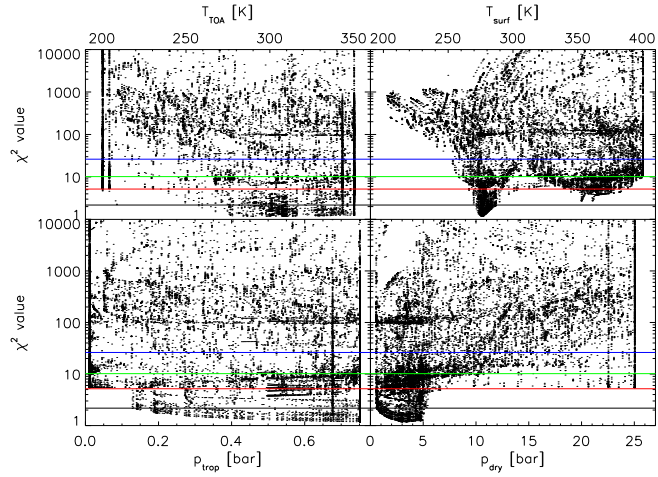}}\\
\caption{$\chi^2$ values as a function of parameter values: temperature profile, for $R$=5, S/N=12 (Darwin). Horizonal lines indicate 1, 2, 3 and 5\,$\sigma$ uncertainties.}
  \label{foundt}
\end{figure}

\begin{figure}[h]
 \resizebox{\hsize}{!}{  \includegraphics*{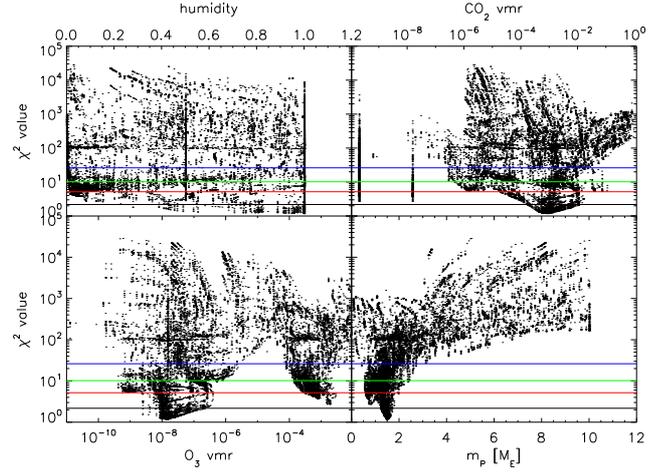}}\\
\caption{$\chi^2$ values as a function of parameter values: composition and mass, for $R$=5, S/N=12 (Darwin). Horizonal lines indicate 1, 2, 3 and 5\,$\sigma$ uncertainties.}
  \label{foundc}
\end{figure}

Figs. \ref{foundt} and \ref{foundc} illustrate how this is done. They show the $\chi^2$ values as a function of the respective parameter values for all points evaluated by MPFIT during the fitting procedure (i.e., not only the best-fit values), a total of about 20,000 points. We chose the $R$=5, S/N=12 case, i.e. the low-resolution Darwin aim (see Table \ref{missions}). Indicated by horizontal lines are the $\Delta \chi^2$=1, $\Delta \chi^2$=4, $\Delta \chi^2$=9 and $\Delta \chi^2$=25 thresholds, corresponding respectively to 1, 2, 3 and 5\,$\sigma$ uncertainties in the estimated parameter value. Also clearly seen are the local minima for $T_{\rm{surf}}$, CO$_2$ and O$_3$. Interestingly, the high O$_3$ concentrations correspond to essentially zero CO$_2$, indicating a strong correlation between both species at low spectral resolution. Furthermore, this also suggests that CO$_2$ cannot be detected in this scenario, since no firm lower limits are found (see below, Table \ref{echofound}). 

As stated in Sect. \ref{fitmodel}, a best-fit model found by a fitting method not necessarily corresponds to a good fit. Therefore, Table \ref{echofound} shows the quality of the fits as indicated by the $p$ value and the $\chi^2_{\rm{red}}$ value. Except for the $R$=5 observations where there are as many parameters as there are data points ($d_f$=0, hence by definition $\chi^2_{\rm{red}}$=$\infty$, Eq. \ref{chi_red}), the $\chi^2_{\rm{red}}$ indicates that good fits are indeed obtained by MPFIT. Also, as stated in Sect. \ref{fitmodel}, MPFIT is dependent on initial conditions. This is indicated in Table \ref{echofound} by the fact that less than 100\,\% of the found best-fits are actually good fits, when adopting a $p$ value of 0.01. However, generally, most estimates of $\vec{x_{\rm{opt}}}$ actually are good fits, even though we used a Newton-type algorithm.

In Table \ref{echofound}, we present the found parameters as well as the 3\,$\sigma$ uncertainties (i.e.,  $\Delta \chi^2$=9, see Figs. \ref{foundt} and \ref{foundc}) for several combinations of spectral resolution and S/N ratio. We use mission specifications as indicated by Table \ref{missions} as well as simulations with increased S/N ratio.

Table \ref{echofound} shows that the temperature profile is largely unconstrained by the observations. This is apparent from the found values of $T_{\rm{TOA}}$ and $p_{\rm{trop}}$ which are, within 3\,$\sigma$, compatible with the boundaries imposed in Table \ref{bound}. Only at $R$=20 with a S/N=20, fit results seem to indicate at least lower limits for $T_{\rm{TOA}}$.

Limits on surface temperatures are tighter, with the fit results in most cases excluding cold surfaces with high confidence. However, for the nominal EChO and Darwin $R$=5 cases, surface temperatures around 260\,K are compatible with the data, which is inconsistent with surface habitability. Even at $R$=20 and S/N=10, surface temperatures above 273\,K are only marginally inferred from the fit. Additionally, for the Darwin $R$=5 and the EChO nominal cases ($R$=5, S/N=12 and $R$=10, S/N=5, respectively), upper limits for $T_{\rm{surf}}$ are consistent with surface temperatures of the order of 400\,K, which is higher than the limit of extremophilic life on Earth (e.g., \citealp{rothschild2001}). Surface pressure is relatively well constrained to within a few bar at $R$=20 and, with an increased S/N of 28, also at $R$=10. For the Darwin $R$=5 cases, however, the entire pressure range of Table \ref{bound} is compatible with the noisy observations. In summary, surface habitability in terms of surface temperature could be inferred in most of the cases, but the temperature profile remains unconstrained.

The planetary mass is generally well-fitted to within a factor of two or better to the true value, indicating that, combined with further constraints based on discovery data, it is probably not a critical parameter in our model. Of course, this is due to the fact that mass correlates with radius (see eq. \ref{mrr}), which then directly influences the emitting surface, hence the number of photons detected. Based on our assumed mass-radius relationship (see Eq. \ref{mrr}), the uncertainties in planetary radius are of the order of 10-50\,\%, which is of the order of current uncertainties in transit surveys.

With respect to atmospheric composition, Table \ref{echofound} implies that the most difficult parameter for the retrieval is actually the humidity $h$. Most cases only allow for weak lower limits, indicating that a significant detection is missing. Also, giving firm upper and lower limits to characterize the humidity is only marginally possible for the $R$=20, S/N=20 case.

The CO$_2$ concentration cannot be constrained for the EChO nominal case, and even a detection is only marginal. At an increased S/N of 26, the EChO resolution allows for CO$_2$ characterization within two orders of magnitude. For the Darwin $R$=5 cases, only upper limits to CO$_2$ could be provided, with lower limits not allowing for a conclusive detection. At $R$=20, in contrast, clear upper limits are found. O$_3$ cannot be characterized accurately with either $R$=5 or $R$=10 nominal S/N ratios, but upon increasing the S/N ratio at the EChO resolution, relatively strong upper limits are found. A spectral resolution of $R$=20 allows for relatively firm upper O$_3$ limits, however at S/N=10 a detection is rather marginal.

These results imply that even with increased S/N ratio (with respect to the current mission design, Table \ref{missions}), both Darwin and EChO are unlikely to be able to provide accurate constraints on atmospheric composition and temperature structure, even though surface temperature and, to some extent, surface pressure are in principle accessible, confirming results from Sects. \ref{sfccond} and \ref{atmoret}.

\begin{figure*}
  \includegraphics{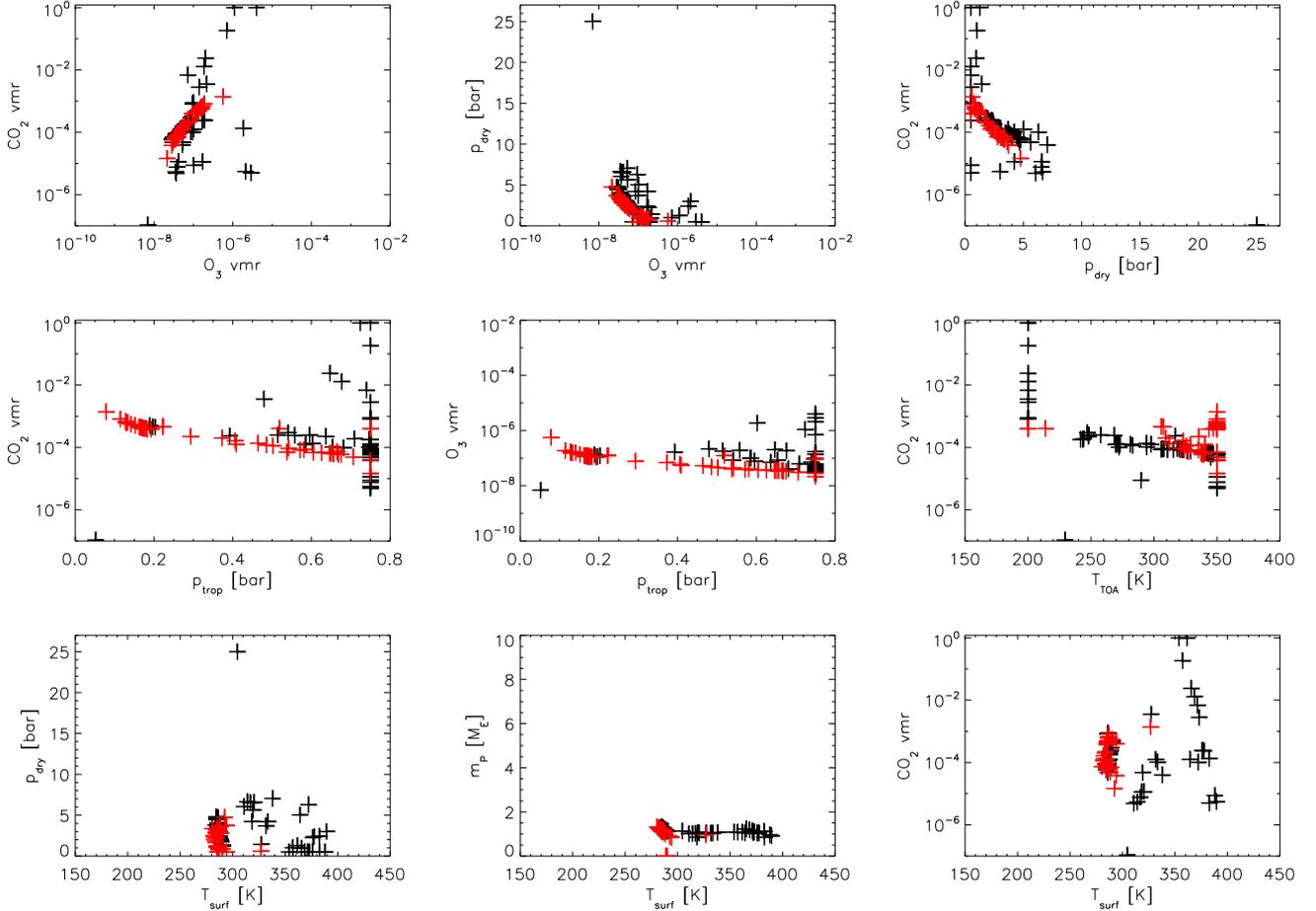}\\
\caption{Scatter plots of best-fit parameters from 100 randomly initialized fits. Black: EChO case, $R$=10, S/N=6, red: Darwin medium-resolution case, $R$=20, S/N=10. }
  \label{corr_echo}
\end{figure*}

In Fig. \ref{corr_echo}, we present scatter plots of the best-fit parameters at both the projected Darwin and EChO designs from Table \ref{missions} to investigate possible correlations between fit parameters. The most obvious correlations are between CO$_2$, O$_3$ and $p_{\rm{dry}}$ (upper panels). CO$_2$ and O$_3$ concentrations are generally lower at higher surface pressures. This is related to the depth of the 15\,$\mu$m and 9.6\,$\mu$m bands, respectively. The apparent correlation between CO$_2$ and O$_3$ is interesting in regards of the discussion on possible false-positive detections of both species (see also discussion below). Correlations between CO$_2$ and O$_3$ with $T_{\rm{TOA}}$ and $p_{\rm{trop}}$ become apparent (middle panels), at least at better S/N ratio and higher spectral resolution (red crosses in Fig. \ref{corr_echo}). At higher values of $p_{\rm{trop}}$, CO$_2$ and O$_3$ concentrations are systematically lower. This is related to the parametrization of the model stratosphere. Both species are best probed in bands originating in the lower stratosphere. Since higher $p_{\rm{trop}}$ values mean warmer stratospheres (all other parameters being equal, see e.g. eq. \ref{upper_atmo} and Fig. \ref{t_illu}), higher $p_{\rm{trop}}$ therefore translates into lower concentrations of CO$_2$ and O$_3$. An analogous argument is valid for the apparent weak correlation between CO$_2$ and O$_3$ with $T_{\rm{TOA}}$. As stated above, there also is a weak correlation between planetary mass and $T_{\rm{surf}}$. However, $T_{\rm{surf}}$ does not show any strong correlation with surface pressure or atmospheric composition (bottom panels).

\section{Discussion}

\label{discussion}

\subsection{Atmospheric model}

In reality, a full atmospheric model which is able to reproduce observations of terrestrial, potentially habitable planets should contain many more parameters than the one presented here (see Sect. \ref{theatmosphere}).

One important factor is the presence of clouds in the atmosphere which have been omitted in the present work (see Sect. \ref{theatmosphere}). Clouds could alter the atmospheric temperature profiles and have a potentially large influence on surface temperatures (e.g., \citealp{forget1997}, \citealp{kitzmann2010}). Furthermore, they also affect the spectral appearance of a planet in emission spectra (e.g., \citealp{tinetti2006synth}). They could mask molecular absorption bands and inhibit the probing of the lower atmosphere and surface (hence, the retrieval of surface conditions). On Earth, many different types of clouds exist in terms of, e.g., vertical and horizontal cloud extension, optical properties or composition. Cloud cover is highly variable both spatially and temporarily. This adds many complications both for atmospheric modeling and spectral retrieval since usually, many single observations must be added to obtain reasonable S/N ratios. 

In addition, many more radiatively active species besides the ones considered in this work could have an impact on temperature structure and emission spectra. In particular, the biomarker species nitrous oxide could be detectable at least in high-resolution spectra given specific planetary scenarios (e.g., early Earth, \citealp{grenfell2011}). SO$_{2}$ and other sulphur species have been proposed in anoxic atmospheres or atmospheres with high volcanic activity (e.g., \citealp{kaltenegger2010so2}, \citealp{domagal2011}). H$_{2}$, given appropriate concentrations of larger than $\approx$10$^{-3}$,  could also be detectable due to distinct collision-induced absorption  (e.g., \citealp{borysow1991h2h2_fund_low}). Additionally, the impact of the stellar radiation on the temperature profile should be taken into account since variations even for fixed atmospheric composition are potentially large (e.g., \citealp{kitzmann2010}). 

Furthermore, many photochemical studies showed that the influence of central star on atmospheric composition cannot be neglected. As an example, the biomarker-relevant gases methane and chloromethane could build up to large concentrations given a suitable stellar UV radiation environment such as on planets around M dwarf stars (e.g., \citealp{Seg2005}). The CO concentration was shown to increase for planets orbiting M dwarfs (e.g., \citealp{rauer2011}). However, using (photo-)chemical models introduces many more additional parameters, e.g., number of important species for catalytic cycles, planetary surface fluxes, stellar and orbital parameters, geochemical cycles, etc., which would complicate the retrieval of planetary properties. Note, however, that constraining CO$_2$ concentrations to within a few orders of magnitude might be indicative of the existence of a silicate weathering cycle which controls CO$_2$ \citep{abbot2012}. Still, the potential importance of using consistent modeling tools is demonstrated by, e.g., \citet{selsis2000} or \citet{rauer2011}. Coupling temperature structure, stellar radiation and atmospheric photochemistry could cause biosignatures such as the O$_3$ band to disappear in the emission spectra.

Planetary and orbital properties such as rotation rate, obliquity and eccentricity induce diurnal and seasonal variability in the emission and reflectance spectra, which again poses a problem for the retrieval, both because of additional free parameters and, in the case of thermal spectra, due to severe degeneracies between obliquity and thermal inertia (e.g., \citealp{gaidos2004}, \citealp{cowan2012}). Surface properties, e.g., oceans, continents, vegetation, ice and snow cover, also influence planetary spectra (e.g., \citealp{hu2012}). Numerous studies have shown how to retrieve such information from spectrophotometry of Earth in the visible spectral range (e.g., \citealp{oakley2009}, \citealp{fujii2010}, \citealp{livengood2011}). However, in a full planetary model such properties would increase the number of free parameters, hence render retrieval more complex and most likely prohibit conclusions on habitability.

The bulk properties of a planet, e.g. its mass-radius relationship, does not necessarily follow the one obtained for rocky, terrestrial planets used here \citep{sotin2007}. The masses and radii of super-Earths discovered so far span a large range (e.g., \citealp{leger2009}, \citealp{charb2009}, \citealp{lissauer2011}, \citealp{batalha2011}). Assuming different bulk compositions for the planet then leads to very different mass-radius relationships (e.g., \citealp{fortney2007}).

Regarding the parametric nature of our model, note that self-consistent modeling could provide a better constraint on CO$_2$ and, indirectly, on O$_3$, since detecting O$_3$ has a lot to do with constraining CO$_2$ (see discussion below). Therefore, the retrieval results presented here might be conservative. Future studies could use real observations of Earth or Mars, or results from consistent modeling studies, as target scenarios and investigate whether accurate and correct retrieval is possible in such more realistic cases.

\subsection{Observational constraints}

In order to reduce the number of fit parameters and break existing degeneracies in emission spectra, observational constraints and additional information must be used when interpreting the spectra. Many important planetary properties could be determined beforehand, such as mass and radius.  Measurements of the planetary mass are available through a number of methods, such as astrometry (e.g., \citealp{malbet2011}), radial velocity, transit timing variations in multiple systems (e.g., \citealp{lissauer2011}) or dynamical models (e.g., \citealp{beust2008}, \citealp{mayor2009gliese}). The planetary radius could be estimated for transiting planets. 

In addition, for transiting planets it is possible to combine emission spectroscopy with transmission spectroscopy. Transmission spectroscopy is, to first order, sensitive to atmospheric scale height, hence mean atmospheric weight and planetary gravity, and only weakly dependent on the vertical temperature structure. Thus, degeneracies between temperature and atmospheric composition could be broken (see, e.g., \citealp{benneke2012}). Due to the large spectral range covered by EChO, such transmission observations could be performed with the same mission as the emission spectroscopy, in contrast to Darwin which would need to rely on other observation facilities. Another complementary method, especially on non-transiting planets, could be to use phase curves and the variation spectrum (e.g., \citealp{harrington2006}, \citealp{seager2009}, \citealp{selsis2011}, \citealp{maurin2012}), a technique which could be used by the Darwin misson.

Broadband optical photometry (over a spectral range of about 0.2-2\,$\mu$m) could be used to measure the reflected light $I_R$ of the planet, which would impose constraints on the Bond albedo $A_B$ and the planetary radius ($I_R\sim A_B\cdot r_P^2$). Since $A_B$ depends strongly on atmospheric composition and surface pressure through absorption of stellar radiation and Rayleigh scattering and Eq. \ref{mrr} relates mass and radius, such measurements could in principle provide useful additional constraints. However, the contrast in reflected light between star and planet is of the order of 10$^{-8}$-10$^{-10}$ for terrestrial, habitable-zone planets. Hence, observations would be very challenging and, consequently, constraints actually may be rather weak.

The orbital period of the planet is determined from the discovery data rather precisely. The orbital distance is determined from the orbital period based on stellar mass (and/or radius, for transiting planets). Combining orbital distance with the stellar energy distribution obtained either from spectroscopy or stellar model atmospheres (e.g., \citealp{buser1992} or \citealp{hauschildt1999}), it is possible to compute the stellar energy input on the planet. Then, assuming global energy conservation for the planet, it would be possible to eliminate model scenarios which would be prohibited, thus possibly limiting the parameter space.

\subsection{Real observations}

In practice, many of the aforementioned potential limits on parameters are rather difficult to obtain. For example, the detection and verification of a planetary signal is very challenging for low-mass planets (see, e.g., \citealp{ferraz2011}, \citealp{pont2011}, \citealp{hatzes2011} for the CoRoT-7 system, \citealp{vogt2010gliese}, \citealp{tuomi2011}, \citealp{gregory2011} for the GL\,581 system). Also, uncertainties on stellar parameters, usually derived from stellar models, are the main problem for derived planetary parameters such as radius and mass, as is for example the case for GJ\,1214\,b (e.g., \citealp{carter2011},  \citealp{sada2010}, \citealp{berta2011}) where the estimated planetary radius changes by 15\,\%, depending on the adopted stellar model.

Other practical concerns are instrumental constraints such as available filters, wavelength coverage and spectral resolution of the telescopes used for the observations or large time spans between observations. Constraints obtained with the same mission (or instrument, even) are usually preferrable to constraints obtained with different observational setups due to systematic noise.

\subsection{Biomarker and bioindicators}

In this work, we considered ozone as a biomarker, since on Earth it is mainly produced from oxygen photochemistry, and the oxygen is provided by the biosphere.

However, interpreting the presence of ozone as a bioindicator is not necessarily straightforward. In other words, a 2\,$\sigma$ detection of ozone is not equivalent to a 2\,$\sigma$ detection of life. Many studies have shown that ozone can be produced abiotically (e.g.,  \citealp{selsis2002}, \citealp{segura2007}, \citealp{domagal2010}). Additionally, it has been shown that ozone persists in the atmospheres of Earth-like planets even at drastically reduced oxygen concentrations and that ozone concentrations depend crucially on the UV radiation field of the central star (e.g., \citealp{Seg2003}, \citealp{Grenf2007asbio}). Very thin ozone layers have indeed been detected both in the atmospheres of Venus and Mars (e.g., \citealp{lebonnois2006}, \citealp{montmessin2011}), although concentrations are far too low to being detectable on exoplanets.

The absence of ozone inferred from observations, however, does not imply that ozone is not present in the atmosphere, as a masking of absorption features is certainly possible, as shown by, e.g., \citet{schindler2000}, \citet{selsis2002} or \citet{vparis2011}.

The most promising way of identifying biospheric signatures in atmospheric spectra seems therefore the simultaneous  presence of species such as water, ozone, oxygen and methane, as proposed by, e.g., \citet{sagan1993} or \citet{selsis2002}.

\section{Conclusions}

\label{concl}

In this work, we investigated the possibility of the retrieval of planetary properties (such as atmospheric composition, temperature structure, radius) from emission spectra. We used a parametric atmosphere model to create arbitrary atmospheres of hypothetical habitable planets. The atmospheric profiles were then used to calculate synthetic, noisy observations. Based on these observations, we tried to fit the model parameters and investigate potential degeneracies.

We have used two design concepts (in terms of S/N ratio and spectral resolution) of exoplanet space missions. Results imply that surface conditions can be characterized relatively well (to within $\sim$10\,K at 3\,$\sigma$) with S/N ratios between 10-30, depending on spectral resolution. This would allow for an assessment of the potential habitability of the target planet. However, even with high S/N ratios of the order of 20 or higher, emission spectra did not allow for accurate determination of atmospheric composition. The biosignature of O$_3$ as well as the atmospheric H$_2$O content are not well constrained with current mission designs. The temperature structure could not be retrieved.

Since achieving such high S/N ratios (or even higher values) is rather difficult in the foreseeable future, single-visit emission spectroscopy alone is most likely not capable of characterizing the atmospheres of potentially habitable planets. In order to obtain higher accuracy on retrieved parameters of habitable planets, emission spectroscopy must be combined with other techniques, e.g. transmission spectroscopy, phase curves or photometry in the visible spectral range. An investigation of such possibilities, and whether this would allow for atmospheric characterization, is subject of future studies. 

\begin{acknowledgements}

P. von Paris, P. Hedelt and F. Selsis acknowledge support from the European Research
Council (Starting Grant 209622: E$_3$ARTHs). This research has been supported in parts by the Helmholtz Association
through the research alliance "Planetary Evolution and Life". Stimulating discussions with H. Rauer and Sz. Csizmadia are gratefully acknowledged. We further acknowledge helpful comments by the referee. We thank A. Borysow for making freely available the Fortran programs to calculate the collision-induced absorption coefficients for N$_2$.

\end{acknowledgements}

\bibliographystyle{aa}
\bibliography{literatur_retrieval}

\end{document}